\documentclass[11pt,a4paper]{article}
\pdfoutput=1
\usepackage{jheppub}
\usepackage{graphicx}
\usepackage{lipsum}
\usepackage{cancel}
\usepackage{slashed}
\usepackage{soul}
\usepackage{mathtools}
\usepackage{bm}
\usepackage{float}
\usepackage{caption}
\usepackage[all]{hypcap}
\usepackage{tikz}
\usepackage{tikz-feynman}
\usepackage{pgfplots}
\usepackage{bbm}
\usepackage{xcolor}
\usepackage{color, colortbl}
\usepackage{subcaption}
\usepackage[font=footnotesize,labelfont=bf]{caption}

\usepackage{amsmath}

\allowdisplaybreaks

\def\det{\mathrm{det}}
\def\Z{\mathbb{Z}}
\def\R{\mathbb{R}}
\def\H{\mathcal{H}}
\def\M{\mathcal{M}}
\def\C{\mathbb{C}}
\def\cC{\mathcal{C}}

\def\Tr{\mathrm{Tr~}}

\usepackage{xspace}

\newcommand{\be}{\begin{equation}}
\newcommand{\ee}{\end{equation}}
\newcommand{\bea}{\begin{eqnarray}}
\newcommand{\eea}{\end{eqnarray}}
\newcommand{\beq}{\begin{equation}}
\newcommand{\eeq}{\end{equation}}
\newcommand{\no}{\nonumber}

\newcommand{\cL}{{\cal L}}
\newcommand{\cM}{{\cal M}}
\newcommand{\cO}{{\cal O}}

\newcommand{\colordiagram}[1]{
\begin{subfigure}[b]{0.31\textwidth}
\begin{tikzpicture}
\begin{feynman}
#1
\end{feynman}
\end{tikzpicture}
\end{subfigure}
}

\def\fermions{
    \vertex [] (psi1) {\footnotesize $\overline{\Psi}_L$};
    \vertex [right=0.8in of psi1] (c0);
    \vertex [right=1.6in of psi1] (psi2) {\footnotesize $\Psi_R$};
    \vertex [above left =0.05in and 0.05in of c0.east] (L0);
    \vertex [above right =0.05in and 0.05in of c0.west] (R0);
    \vertex [above  =0.05in of psi1.east] (psi1V);
    \draw[red]  (L0)--(psi1V);
    \vertex [above  =0.05in of psi2.west] (psi2V);
    \draw[blue,dash pattern={on 3pt off 1pt}]  (psi2V)--(R0);
    \diagram* {
    (psi1) -- [fermion] (c0)  -- [anti fermion] (psi2)
    };
}

\def\sep{0.35in}

\newcommand{\HiggsII}[2]{
    \vertex [above=\sep of c0] (c1);
    \vertex [above=\sep of c1] (c2) {\footnotesize $\H^{22}_a$};
    \diagram* {(c0)--[scalar] (c1)--[scalar] (c2)};
    \vertex [ left  =0.05in of c2.south] (#1);
    \vertex [ right  =0.05in of c2.south] (#2);
}
\newcommand{\HiggsIII}[2]{
    \vertex [above=\sep of c0] (c1);
    \vertex [above=\sep of c1] (c2);
    \vertex [above=\sep of c2] (c3) {\footnotesize $\H^{22}_a$};
    \diagram* {(c0)--[scalar] (c1)--[scalar] (c2)--[scalar] (c3)};
    \vertex [ left  =0.05in of c3.south] (#1);
    \vertex [ right  =0.05in of c3.south] (#2);
}

\newcommand{\phiLI}[1]{

    \vertex [left=0 in of #1] (A);

    \vertex [below left =0.02in and 0.05in of A.west] (L1);

    \vertex [above left =0.02in and 0.05in of A.west] (L2);

    \vertex [left=0.3in of A] (E) {\footnotesize $\langle\Phi_L\rangle$};

    \diagram* {(A) --[scalar] (E)};

    \vertex [left=0.27in of L1] (B);

    \draw[red] (B)--(L1);

    \vertex [left=0.27in of L2] (D);

    \draw[red] (L2)-- (D);

}

\newcommand{\phiRI}[1]{

    \vertex [right=0 in of #1] (A);

    \vertex [below right =0.02in and 0.05in of A.east] (R1);

    \vertex [above right =0.02in and 0.05in of A.east] (R2);

    \vertex [right=0.3in of A] (E) {\footnotesize $\langle\Phi_R\rangle$};

    \diagram* {(A) --[scalar] (E)};

    \vertex [right=0.27in of R1] (B);

    \draw[blue,dash pattern={on 3pt off 1pt}] (B)--(R1);

    \vertex [right=0.27in of R2] (D);

    \draw[blue,dash pattern={on 3pt off 1pt}] (D)-- (R2);

}

\newcommand{\LII}[2]{
\draw[red,postaction={decoration={markings,mark=at position 0.5 with {\arrow{<<}}},decorate}]  (#1)-- (#2);
}
\newcommand{\LI}[2]{
\draw[red,postaction={decoration={markings,mark=at position 0.5 with {\arrow{<}}},decorate}]  (#1)-- (#2);
}

\newcommand{\LrII}[2]{
\draw[red,postaction={decoration={markings,mark=at position 0.5 with {\arrow{>>}}},decorate}]  (#1)-- (#2);
}
\newcommand{\LrI}[2]{
\draw[red,postaction={decoration={markings,mark=at position 0.5 with {\arrow{>}}},decorate}]  (#1)-- (#2);
}

\newcommand{\RII}[2]{
\draw[blue,dash pattern={on 3pt off 1pt},postaction={decoration={markings,mark=at position 0.5 with {\arrow{>>}}},decorate}]  (#1)-- (#2);
}
\newcommand{\RI}[2]{
\draw[blue,dash pattern={on 3pt off 1pt},postaction={decoration={markings,mark=at position 0.5 with {\arrow{>}}},decorate}]  (#1)-- (#2);
}

\begin{document}

\title{Electroweak-flavour and quark-lepton unification: \\
a family non-universal path}

\author{Joe Davighi,}
\author{Gino Isidori,}
\author{and Marko Pesut}

\affiliation{Physik-Institut, Universit\"at Z\"urich, CH 8057 Z\"urich, Switzerland}
\emailAdd{joe.davighi@physik.uzh.ch}
\emailAdd{isidori@physik.uzh.ch}
\emailAdd{marko.pesut@physik.uzh.ch}

\abstract{We present a  family-non-universal  extension of the Standard Model where the 
the first two families feature both quark-lepton and electroweak-flavour unification,
via the  $SU(4) \times Sp(4)_L \times Sp(4)_R$ gauge group, 
whereas quark-lepton unification for the third family is realised \`a la Pati-Salam.
Via staggered symmetry breaking steps, this construction offers a natural explanation for the observed hierarchical 
pattern of fermion masses and mixings, while providing a natural suppression for 
 flavour-changing processes involving the first two generations. The last-but-one step in the 
 symmetry-breaking chain is a non-universal 4321 model, characterised by 
 a vector leptoquark naturally coupled mainly to the third generation. 
 The stability of the Higgs sector points to a 4321$\to$SM symmetry-breaking scale 
 around the TeV, with interesting phenomenological consequences in $B$ physics 
 and collider processes that differ from those of other known 4321  completions. 
}

\maketitle

\section{Introduction}

The Standard Model (SM) of particle physics is currently the most accurate theoretical framework to describe microscopic phenomena. 
The SM successfully passed several precision tests, and no new degrees of freedom have emerged yet by the direct exploration of the TeV energy domain at the LHC. Nevertheless, the SM is plagued by a significant number of open issues. The two we deal with in this paper, 
namely the \emph{flavour puzzle} and the possible \emph{unification} of quarks and leptons, are related to the matter content of the SM.
The flavour puzzle refers to the highly non-generic pattern of masses and mixings of the three families of quarks and leptons, which has no justification within the SM.  Equally puzzling is the peculiar assignment of the $U(1)_Y$ charges for the SM fermions that, despite being justified \emph{a posteriori} by the requirement of anomaly cancellation, naturally points toward some form of unification of quark and lepton quantum numbers  at high energies.

Attempts to provide dynamical justifications of the flavour puzzle, and attempts to unify quarks and leptons into representations 
of new (non-Abelian) gauge groups, both have a long history. However, to a large extent, these two efforts proceeded in parallel until recently.
Unification was pursued in a flavour-blind manner, extending the SM gauge group into a grand-unified group acting in the same way for the three 
fermion generations~\cite{Georgi:1974sy}. The flavour problem was addressed via appropriate  horizontal symmetries (global or gauged, continuous or discrete), commuting with the aforementioned  unified gauge symmetry, as proposed for instance in \cite{Froggatt:1978nt}.
The factorisation of flavour and gauge symmetries was phenomenologically motivated by the universality of the SM gauge group and, to some extent, by simplicity. But it is certainly not the only option. The path we explore in this paper is a different one: it is based on the assumption that the gauge group in the ultraviolet (UV) is fundamentally family non-universal.

In the case of family non-universal gauge groups, the flavour problem is addressed via  a cascade of symmetry breaking steps
occurring at different energy scales, from the initial non-universal group to more universal ones, eventually ending 
with the SM, as in~\cite{Dvali:2000ha,Panico:2016ull,Bordone:2017bld,Allwicher:2020esa,Barbieri:2021wrc,Fuentes-Martin:2022xnb}. 
Qualitatively, the light families are generated at some high scale, where the non-universality of the light families becomes manifest ({\em i.e.}~at the scale where the light fermions have new, non-universal, dynamical  interactions). This implies a suppression with respect to
the third generation Yukawa couplings, which are generated at a lower scale.  This type of construction potentially addresses also another key open issue of the SM: a (stable) separation of the scale stabilising the Higgs sector and the scale of new dynamics affecting the light families.  With such separation,  the new dynamics stabilising the Higgs sector, which necessarily couples strongly also to the third family, can still be quite close to the TeV scale while avoiding (at least in part) the tight constraints derived from processes involving the light families.

In addition to these general  (top-down) arguments, interest in family non-universal gauge groups has arisen recently from a pure bottom-up
perspective because of the $B$-physics anomalies, {\em i.e.}~the deviations from the SM predictions observed in semileptonic $B$-meson decays (see~\cite{London:2021lfn} and references therein).  An interesting hypothesis describing well all present data is the extension of the SM field content by a massive vector leptoquark field ($U_1$),  in the TeV mass range, coupled mainly to the third generation~\cite{Buttazzo:2017ixm}.  
The $U_1$ field has the right couplings and quantum numbers to be the broken generator of a (non-universal) $SU(4)_3$ group
acting on the third family.  This, in turn, has led to the so-called 4321 models: a construction based on the (TeV-scale) gauge symmetry
 $SU(4)_3\times SU(3)_l \times SU(2)_L \times U(1)_X$~\cite{DiLuzio:2017vat,Bordone:2017bld,Greljo:2018tuh}, where $SU(3)_l$ acts only on the light families and color is the family diagonal subgroup of $SU(4)_3\times SU(3)_l$.
Besides offering a successful description of the $B$-physics anomalies, this set-up features~i) quark and lepton unification 
\`a la Pati-Salam~\cite{Pati:1974yy} for the third generation, and ii)~an accidental $U(2)^5$ global flavour symmetry acting on the light families. 
The latter is known to be an excellent first-order approximation to the SM Yukawa couplings and, if broken in a minimal way, a key ingredient to ensuring sufficient protection against flavour-violating effects involving the light families~\cite{Barbieri:2011ci,Isidori:2012ts}.

It is natural to try to merge the indication of a 4321 gauge group at the TeV scale with the more general 
top-down considerations about non-universal gauge groups presented above. 
From this perspective, the 4321 group should be viewed as the last-but-one step of the symmetry breaking chain from the UV theory down to the SM. 
Attempts of this type have been presented in~\cite{Bordone:2017bld,Fuentes-Martin:2020bnh,Fuentes-Martin:2022xnb,FernandezNavarro:2022gst}. In this paper we present a new proposal along this line, which is significantly different from all previous attempts.

The novel aspect of our construction is the concept of \emph{electroweak-flavour unification}, which invokes an $Sp(2N_f)_L \times Sp(2N_f)_R$ gauge symmetry (in the general case of $N_f$ families), recently proposed in~\cite{Davighi:2022fer}.   Rather than imposing this structure to unify all three fermion families, 
as suggested in~\cite{Davighi:2022fer}, we here propose a different path for light {\em vs.}~heavy families: both sectors feature quark-lepton 
unification via independent $SU(4)$ groups, while only the light sector features electroweak-flavour unification via $Sp(4)_L \times Sp(4)_R$. 
With this proposal the separation between the third family and the other two is  ``hard-coded''  in the UV, where the gauge symmetry is the direct product of groups acting separately on the different fermion sectors. This is a good premise to justify in a natural way the large mixing occurring in the light family sector, controlled by the Cabibbo angle ($|V_{us}|\sim 0.2$), {\em vs.}~the small heavy-light mixing controlled by $|V_{ts}|\sim 0.04$. 
As we shall see, assuming a sufficiently high-scale for the $Sp(2N_f)_L \times Sp(2N_f)_R$ symmetry breaking, this set-up 
also provides a natural justification for the minimal breaking of the  $U(2)^5$ flavour symmetry~\cite{Barbieri:2011ci}, 
which ensures an efficient suppression of non-SM effects in flavour-changing processes involving the light families.\footnote{Arguably, a $U(2)^5$ flavour symmetry appears even more motivated after the recent $R_{K^{(\ast)}}$ measurement by the LHCb collaboration~\cite{LHCb:2022zom}, which
provides a further strong constraint on non-universality between the two light lepton generations. }

The mixing between the two sectors happens via scalar fields charged under both gauge groups, and appropriate vector-like fermions. 
While the observed flavour hierarchies (masses and mixings) fix only the ratios between the different scales of this construction, an absolute indication of the new energy scales comes from the stability of the field combination responsible for electroweak symmetry breaking, {\em i.e.}~the effective SM Higgs sector. The latter requires the lowest non-standard symmetry breaking step (4321$\to$SM) to occur around the TeV scale. This is a further indication for new dynamics coupled mainly to the third generation in this energy domain, independent from that obtained from the $B$-physics anomaly.  As we shall show, in this framework the data-theory comparison in $B$-physics improves, but the low-energy signatures of the model are different from those derived in the other known 4321 completions. This fact, together with the specific TeV-scale phenomenology of the model, could help in the future to identify this motivated scenario. 

The paper is organised as follows. In Section \ref{genintro} we introduce the UV gauge group as well as the matter content of the model. We discuss the symmetry breaking pattern and briefly review some basic facts about $Sp(4)$ Lie groups. In Section \ref{dyngenyuk} we present a detailed derivation of the 
Yukawa couplings for all the SM fermions. Section~\ref{sec:Higgs-stability} is devoted to the last step of the symmetry breaking chain, which allow us to anchor the various scales of the model from the stability of the Higgs sector. 
In this Section we also derive predictions for both $B$-physics and collider phenomenology.
Finally, we conclude and discuss some interesting future directions in Section \ref{conclu}.

\section{A new flavoured gauge model}\label{genintro}

The model we propose is based on a high-energy gauge group that treats the third family differently from the light two families, which are tightly unified.
The gauge group has the form
\be \label{eq:G}
G= G_{12} \times G_3,
\ee
with the light families charged only under $G_{12}$ and the third family charged only under $G_3$. As described in the Introduction, such a factorization of the gauge sector is strongly-motivated by the observed Yukawa sector alone, together with considerations of naturalness and data from high $p_T$.

Guided by our desire for a third-family aligned $U_1$ leptoquark, we then take 
\be \label{eq:G3}
G_3=SU(4)_3 \times SU(2)_{L,3} \times SU(2)_{R,3}
\ee
to be a Pati--Salam symmetry. For the light families we also unify quarks with leptons \`a la Pati--Salam, while further unifying the electroweak and flavour quantum numbers in the manner recently explored for all three families in Ref.~\cite{Davighi:2022fer}. We thus consider\footnote{
We remark that $\mathfrak{g}=\mathrm{Lie}(G)$ is listed as algebra number 157 in Ref.~\cite{Allanach:2021bfe} (supplementary material), which comprehensively studied semi-simple gauge algebras with possible gauge-flavour unification.}
\be \label{eq:G12}
G_{12} = SU(4)_{1+2} \times Sp(4)_L \times Sp(4)_R,
\ee
where $Sp(4)$ denotes the symplectic group (see \S \ref{sec:conventions}). More minimal choices for $G_{12}$ and $G_3$ could have been made, for example 
$G_3=\mathrm{SM}_3$ and $G_{12} = SU(3)_{1+2} \times Sp(4)_L \times Sp(4)_R \times U(1)_{B-L,1+2}$,
if we do not wish to unify quarks and leptons.
The choices (\ref{eq:G3}) and (\ref{eq:G12}) for $G_3$ and $G_{12}$ are motivated by matter unification in the UV, in particular absorbing $U(1)$ factors in the gauge symmetry, plus the phenomenological interest in vector leptoquarks near the TeV scale.

\subsection{Mathematical notation and conventions} \label{sec:conventions}

In our convention, $Sp(4)\subset SU(4)$ is the 10-dimensional group of $4\times 4$ special unitary matrices $\{U\}$ that moreover satisfy $U^T \Omega U = \Omega$, where the matrix
\be
\Omega=\begin{pmatrix} 0 & \mathbb{I}_2 \\ -\mathbb{I}_2 & 0 \end{pmatrix},
\ee
with $\mathbb{I}_2$ being the 2-by-2 identity matrix. 
The Lie algebra $\mathfrak{sp}(4)$ and its representations are probably familiar to most readers, thanks to the Lie algebra isomorphism $\mathfrak{sp}(4) \cong \mathfrak{so}(5)$. The corresponding Lie group isomorphism is $Sp(4) \cong \mathrm{Spin}(5)$, where $\mathrm{Spin}(5)$ is the double cover of $SO(5)$.

To set out a little more necessary notation, 
we let $\{a_1,a_2,a_3,a_4\}$ denote a basis for the $\C^4$ vector space acted on by the fundamental representation $\bf{4}$ of $SU(4)_{1+2}$, with $\{a_i^\ast\}$ a basis for the conjugate $\bf{\overline{4}}$ representation; correspondingly, let $\{A_i\}$ and $\{A_i^\ast\}$ denote bases for the fundamental and anti-fundamental representations of $SU(4)_3$. We let
$\{b_1, \dots, b_4\}$ and $\{c_1, \dots, c_4\}$ denote bases for the vector spaces (both $\C^4$) acted on by the fundamental of $Sp(4)_L$ and $Sp(4)_R$ respectively, and $\{B_1, B_2\}$, $\{C_1, C_2\}$ denote the bases for the $\C^2$ vector spaces acted on by the fundamental representations of $SU(2)_{L,3}$ and $SU(2)_{R,3}$.

For $Sp(2n)$ groups, like for $SU(2)\cong Sp(2)$, the fundamental and its conjugate representation are isomorphic. It is therefore convenient to introduce a notion of complex conjugation ($\ast$) that, in addition to sending all $\C$-valued components to their ordinary complex conjugates, acts on our basis vectors as $\ast: b_i \mapsto \Omega_{ji} b_j$ and $\ast: c_i \to \Omega_{ji} c_j$. With this definition, the conjugate $\varphi^\ast$ of some field $\varphi$ transforming in the ${\bf 4}$ of $Sp(4)$ also transforms in the ${\bf 4}$.\footnote{
To see this explicitly, let $\varphi = \varphi_i b_i$ denote a field in the fundamental ${\bf 4}$ of $Sp(4)$. Under the $Sp(4)$ action, $\varphi_i \mapsto U_{ij} \varphi_j$, where $U_{ij}$ are the components of a 4-by-4 $Sp(4)$ matrix. The conjugate field $\varphi^\ast$ has components $(\varphi^\ast)_i = \Omega_{ij} \varphi_j^\ast$ with respect to the same basis $\{b_i\}$. Under $Sp(4)$, $(\varphi^\ast)_i \mapsto \Omega_{ij} U_{jk}^\ast \varphi_k^\ast = U_{ij} \Omega_{jk} \varphi_k^\ast = U_{ij} (\varphi^\ast)_j$, where we have used $U^T \Omega U = \Omega$. Thus $(\varphi^\ast)$ and $\varphi$ transform in the same ${\bf 4}$ representation.
} For the $n=1$ case, the factor of $\Omega$ appearing in the automorphism $\ast$ is the familiar factor of $i\sigma_2$ that conventionally appears in complex conjugation of $SU(2)$ doublets.

The $\Omega$ matrices provide invariant tensors for symplectic groups, which we can use to construct singlets. In particular, given two fields $x=x_i b_i$ and $y=y_i b_i$ in the fundamental representation of $Sp(4)$, the contraction
\be \label{eq:sp-contraction}
x_i \Omega^{ij} y_j 
\ee
is an $Sp(4)$ singlet. We will frequently encounter such contractions of $Sp(4)$ fundamentals in the following, where we always suppress writing the indices and the requisite insertions of the $\Omega$ matrix (which are implied). Other than the fundamental, all the representations of $Sp(4)$ that feature in this paper appear in the tensor product of two fundamentals, which is
\be
\bf{4} \otimes \bf{4} = \bf{1} \oplus \bf{5} \oplus \bf{10}\, .
\ee
Here, the $\bf{10}$ is the {\em symmetric} contraction of two fundamentals, which for $Sp(2N)$ Lie groups is isomorphic to the {\em adjoint} representation, and the $\bf{5}$ is the part of the {\em antisymmetric} contraction that remains after subtracting off the singlet (\ref{eq:sp-contraction}).

We also introduce a convenient notation to indicate the `flow' of $Sp(4)$ indices in Feynman diagrams, following~\cite{Davighi:2022fer}, whereby solid red lines (\textcolor{red}{---}) marked with one or two arrows denote specific contractions of $Sp(4)_L$ fundamental representations $x$ and $y$. The number of arrows matches the family of the SM fermions involved in the contraction:
\begin{figure}[H]
\begin{center}
\colordiagram{
	\vertex (x){\footnotesize $y\,] \sim x_1 y_3$};
	\vertex [left=3 cm of x] (y){\footnotesize $[\,x$};
    \LI{x}{y};
}
\colordiagram{
	\vertex (x){\footnotesize $y\,] \sim x_3 y_1$};
	\vertex [left=3 cm of x] (y){\footnotesize $[\,x$};
    \LrI{x}{y};
}
\\
\colordiagram{
	\vertex (x){\footnotesize $y\,] \sim x_2 y_4$};
	\vertex [left=3 cm of x] (y){\footnotesize $[\,x$};
    \LII{x}{y};
}
\colordiagram{
	\vertex (x){\footnotesize $y\,] \sim x_4 y_2$};
	\vertex [left=3 cm of x] (y){\footnotesize $[\,x$};
    \LrII{x}{y};
}
\end{center}
\captionsetup{labelformat=empty}
\caption{}
\addtocounter{figure}{-1}
\end{figure}
\vspace{- 1cm}
\noindent
Similarly, dashed blue lines (\textcolor{blue}{- - -}) will be used to represent $Sp(4)_R$ contractions.

\begin{table}
\begin{center}{\small
\begin{tabular}{|c|c|ccc|ccc|}
\hline
 & Field & $SU(4)_{1+2}$ & $Sp(4)_L$ & $Sp(4)_R$ & $SU(4)_3$ & $SU(2)_{L,3}$ & $SU(2)_{R,3}$ \\
\hline
SM Fermions (chiral) & $\Psi_L^{l}$ & $\bf{4}$ & $\bf{4}$ & $\bf{1}$ & $\bf{1}$ & $\bf{1}$ & $\bf{1}$ \\
    & $\Psi_R^{l}$ & $\bf{4}$ & $\bf{1}$ & $\bf{4}$ & $\bf{1}$ & $\bf{1}$ & $\bf{1}$ \\
    & $\Psi_L^3$ & $\bf{1}$ & $\bf{1}$ & $\bf{1}$ & $\bf{4}$ & $\bf{2}$ & $\bf{1}$ \\
    & $\Psi_R^3$ & $\bf{1}$ & $\bf{1}$ & $\bf{1}$ & $\bf{4}$ & $\bf{1}$ & $\bf{2}$ \\
\hline
\hline
Vector-like fermion & $\Xi_{L/R}$ & $\bf{4}$ & $\bf{1}$ & $\bf{4}$ & $\bf{1}$ & $\bf{1}$ & $\bf{1}$ \\
\hline
\hline
EWSB Higgses & $\H_{1}$ & $\bf{1}$ & $\bf{4}$ & $\bf{4}$ & $\bf{1}$ & $\bf{1}$ & $\bf{1}$ \\
& $\H_{15}$ & $\bf{15}$ & $\bf{4}$ & $\bf{4}$ & $\bf{1}$ & $\bf{1}$ & $\bf{1}$ \\
& $H_{1}$ & $\bf{1}$ & $\bf{1}$ & $\bf{1}$ & $\bf{1}$ & $\bf{2}$ & $\bf{2}$ \\
& $H_{15}$ & $\bf{1}$ & $\bf{1}$ & $\bf{1}$ & $\bf{15}$ & $\bf{2}$ & $\bf{2}$ \\
\hline
\hline
Symmetry breaking scalars & $S_L$ & $\bf{1}$ & $\bf{5}$ & $\bf{1}$ & $\bf{1}$ & $\bf{1}$ & $\bf{1}$ \\
& $S_R$ & $\bf\bar{4}$ & $\bf{1}$ & $\bf{4}$ & $\bf{1}$ & $\bf{1}$ & $\bf{1}$ \\
& $\Phi_L$ & $\bf{1}$ & $\bf{5}$ & $\bf{1}$ & $\bf{1}$ & $\bf{1}$ & $\bf{1}$ \\
& $\Phi_R$ & $\bf{1}$ & $\bf{1}$ & $\bf{5}$ & $\bf{1}$ & $\bf{1}$ & $\bf{1}$ \\
& $\Sigma_L$ & $\bf{1}$ & $\bf{4}$ & $\bf{1}$ & $\bf{1}$ & $\bf{2}$ & $\bf{1}$ \\
& $\Sigma_R$ & $\bf{1}$ & $\bf{1}$ & $\bf{4}$ & $\bf{1}$ & $\bf{1}$ & $\bf{2}$ \\
& $\omega$ & $\bf{4}$ & $\bf{1}$ & $\bf{4}$ & $\bf\bar{4}$ & $\bf{1}$ & $\bf{2}$ \\
\hline
\end{tabular}
}
\end{center}
\caption{Field content of the model. In addition to the SM fermions, there are various scalar fields in which the electroweak symmetry breaking Higgs fields are embedded, as well as other symmetry breaking scalars that break the gauge symmetry down to the SM, plus a vector-like fermion. } \label{tab:Matter_Content}
\end{table}

\subsection{Embedding the Standard Model fields}

The light fermion fields of the SM are all packaged into two 16-component reps, $\Psi_L^{l} \sim ({\bf 4}, {\bf 4}, {\bf 1})\otimes {\bf(1,1,1)}$ which is left-handed and $\Psi_R^{l} \sim ({\bf 4}, {\bf 1}, {\bf 4})\otimes {\bf(1,1,1)}$ which is right-handed, where the label `$l$' stands for `light'. We can represent $\Psi_L^{l}$ and $\Psi_R^{l}$ as $4\times 4$ matrices whose rows transform in the fundamental representation of $Sp(4)_L$ and $Sp(4)_R$ respectively, and whose columns transform in the fundamental of $SU(4)_{1+2}$, {\em viz.}
\begin{equation} \label{eq:fermions}
\Psi_L^{l}=
\begin{pmatrix}
u_{1,L}^r & u_{2,L}^r & d_{1,L}^r & d_{2,L}^r \\
u_{1,L}^g & u_{2,L}^g & d_{1,L}^g & d_{2,L}^g \\
u_{1,L}^b & u_{2,L}^b & d_{1,L}^b & d_{2,L}^b \\
\nu_{1,L} & \nu_{2,L} & e_{1,L} & e_{2,L} \\
\end{pmatrix} \, ,
\qquad
\Psi_R^{l}=
\begin{pmatrix}
u_{1,R}^r & u_{2,R}^r & d_{1,R}^r & d_{2,R}^r \\
u_{1,R}^g & u_{2,R}^g & d_{1,R}^g & d_{2,R}^g \\
u_{1,R}^b & u_{2,R}^b & d_{1,R}^b & d_{2,R}^b \\
\nu_{1,R} & \nu_{2,R} & e_{1,R} & e_{2,R} \\
\end{pmatrix} \, ,
\end{equation}
where $u^a_{i,L}$ denotes a left-handed up quark with colour $a$ and family index $i$, {\em etc.}
With this chiral fermion content, our gauge model is free of both perturbative and non-perturbative gauge anomalies.\footnote{The $SU(4)$ factors have possible perturbative gauge anomalies, but these cancel because there are equal numbers of left- and right-handed Weyl fermions in the (anomalous) fundamental representations of each $SU(4)$. The $Sp(4)$ and $SU(2)$ factors can suffer at most mod 2 anomalies in 4d~\cite{Witten:1982fp}, since these groups only have real and pseudo-real representations. Given that we have even numbers of Weyl fermions charged in the fundamental representation of each $Sp(4)$ and $SU(2)$ factor, all these mod 2 anomalies cancel. Finally, one can rigorously check that there are no further possible non-perturbative anomalies by computing the spin-bordism group $\Omega_5^\text{Spin}\left(B(G_{12} \times G_3)\right) \cong (\Z_2)^4$, using {\em e.g.} the methods of Refs.~\cite{Garcia-Etxebarria:2018ajm,Davighi:2019rcd}.}

There are Higgs fields that couple to both light families, in the representations $\H_1 \sim ({\bf 1}, {\bf 4}, {\bf 4})\otimes {\bf(1,1,1)}$ and $\H_{15} \sim ({\bf 15}, {\bf 4}, {\bf 4})\otimes {\bf(1,1,1)}$, as well as separate Higgs fields  $H_1 \sim  {\bf(1,1,1)}\otimes({\bf 1}, {\bf 2}, {\bf 2})$ and $H_{15} \sim {\bf(1,1,1)}\otimes({\bf 15}, {\bf 2}, {\bf 2})$ that couple to the third family. It will be the latter Higgs fields, $H_1$ and $H_{15}$, that acquire non-zero vacuum expectation values (vevs) which break electroweak symmetry and, via mixing with $\H$ (see \S \ref{sec:Higgs-stability}), mediate the light Yukawa couplings.

The representations of all SM fields are recorded in Table~\ref{tab:Matter_Content}, along with all the extra fields (many scalars and one vector-like fermion) that will feature in the model.

\subsection{Fundamental Yukawa interactions} \label{sec:dim4yuks}

The renormalisable Yukawa couplings between the SM fermions and these various Higgs fields are
\begin{align}
-\mathcal{L} \supset &\sum_{a\in\{1,15\}}\left[\, y_a^{l}\, \Tr (\overline{\Psi}_L^{l} \H_a \Psi_R^{l}) + \bar{y}_a^{l}  \,\Tr (\overline{\Psi}_L^{l} \H_{a}^{\ast} \Psi_R^{l}) \,\right] \nonumber \\
+&\sum_{a\in\{1,15\}} \left[\, y_a^3\,\Tr (\overline{\Psi}_L^{3} H_a \Psi_R^{3}) + \bar{y}_a^{3}  \,\Tr (\overline{\Psi}_L^{3} H_{a}^{\ast} \Psi_R^{3}) \,\right] + \text{ h.c. } \, 
\end{align}
The terms in the first line couple the light fermions $\Psi_{L,R}^l$ to the $\H_a$ Higgs fields. 
The terms in the second line couple the third family fermions $\Psi_{L,R}^3$ to the $H_a$ Higgs fields.

The model will also feature a vector-like fermion (VLF) in the representation $\Xi \sim {\bf (4,1,4)}\otimes {\bf(1,1,1)}$, charged only under the light-flavour group -- specifically, in the same representation as $\Psi_R^{l}$. This VLF is needed to introduce mixing between the third family SM fermions and the light generations, as discussed in \S \ref{sect:Yuk_mix}.
There are additional renormalisable Yukawa interactions involving $\Xi$, some with the $\mathcal{H}_a$ Higgs fields, and others with the scalar field $\omega \sim {\bf (4,1,4) } \otimes {\bf (\bar{4},1,2) }$ (see Table~\ref{tab:Matter_Content}) that will play a role in breaking the UV gauge symmetry down to the SM. These extra Yukawa interactions are
\begin{align} \label{eq:Xi_yukawa}
-\mathcal{L} \supset \,
&\lambda\, \Tr (\overline{\Xi}_L \omega \Psi_R^3 ) 
+ \bar{\lambda}\, \Tr (\overline{\Xi}_L \omega^{\ast} \Psi_R^3 ) \\
+\sum_{a\in\{1,15\}} &\left[\, \kappa_a \Tr (\overline{\Psi}_L^{l} \H_a \Xi_R) + \bar{\kappa}_a  \,\Tr (\overline{\Psi}_L^{l} \H_{a}^{\ast} \Xi_R) \,\right] +\text{~h.c.} \nonumber
\end{align}
All the coefficients of these fundamental Yukawa interactions, namely 
\be
\{y_a^l, \bar{y}_a^l, y_a^3, \bar{y}_a^3, \lambda, \bar{\lambda}, \kappa_a, \bar{\kappa}_a\},
\ee 
are presumed to be independent $\mathcal{O}(1)$ numbers.\footnote{The term in (\ref{eq:Xi_yukawa}) with coupling $\bar{\lambda}$ will not in fact appear in the final formulae we obtain for the physical fermion mixing angles -- but we include it here for completeness.
}

\begin{figure} [h]
\begin{center}
\begin{tikzpicture}
\node at (0,9){$\textcolor{blue}{\left[SU(4)_{1+2}\times Sp(4)_L \times Sp(4)_R\right]} \times \left[SU(4)_3\times SU(2)_{L,3} \times SU(2)_{R,3}\right]$};
\node at (-3,7.5){$\textcolor{blue}{{\scriptstyle SU(3)_{1+2}\times SU(2)_{L,1} \times SU(2)_{L,2} \times SU(2)_{R,1} \times U(1)_{R}^{\prime\prime}}}$};
\node at (0,6){$\textcolor{blue}{\left[SU(3)_{1+2}\times SU(2)_{L,1+2} \times U(1)_{Y,1+2}\right]} \times \left[SU(4)_3\times SU(2)_{L,3} \times SU(2)_{R,3}\right]$};
\node at (0,4){$\textcolor{red}{SU(4)_3 \times SU(3)_{1+2} \times SU(2)_L \times U(1)_R^\prime}$};
\node at (0,2){$\textcolor{red}{SU(3)\times SU(2)_L\times U(1)_Y}$};
\node at (-7.3,8){$\textcolor{blue}{\Lambda_{12}}$};
\node at (-7.3,7){$\textcolor{blue}{\epsilon\Lambda_{12}}$};
\node at (-7.3,5){$\Lambda_{\Sigma}$};
\node at (-7.3,3){$\textcolor{red}{\Lambda_{4321}}$};
\draw[->, draw = blue] (-3,8.6)--(-3,7.7);
\draw[->, draw = blue] (-3,7.3)--(-3,6.4);
\node[anchor=east] at (-3,8){$\textcolor{blue}{{\scriptstyle\langle S_L \rangle}}$};
\node[anchor=east] at (-2,8){$\textcolor{blue}{{\scriptstyle\langle S_R \rangle}}$};
\draw[->,dashed] (3,8.6)--(3,6.4);
\node[anchor=east] at (-3,7){$\textcolor{blue}{{\scriptstyle\langle \phi_L \rangle}}$};
\node[anchor=east] at (-2,7){$\textcolor{blue}{{\scriptstyle\langle \phi_R \rangle}}$};
\draw[->] (0,5.6)--(0,4.4);
\node[anchor=east] at (-0.5,5){$\langle\Sigma_L\rangle$};
\node[anchor=west] at (0.5,5){$\langle\Sigma_R\rangle$};
\draw[->, draw = red] (0,3.6)--(0,2.4);
\node[anchor=east] at (-0.5,3){$\textcolor{red}{\langle \omega \rangle}$};
\end{tikzpicture}
\end{center}
\caption{The symmetry breaking scheme in our model.
The portion of the diagram coloured \textcolor{blue}{blue} is responsible for generating the light family Yukawas, and thus for controlling the flavour structure of the light family sector. The portion coloured \textcolor{red}{red} mimics the low-energy breaking of so-called `4321 models', delivering in particular a $U_1$ leptoquark coupled mostly to the third family. We will eventually suggest concrete scales for each $\Lambda$ that appears on the vertical `axis' -- see Fig.~\ref{Fig:scales}. \label{fig:SSB} }
\end{figure}
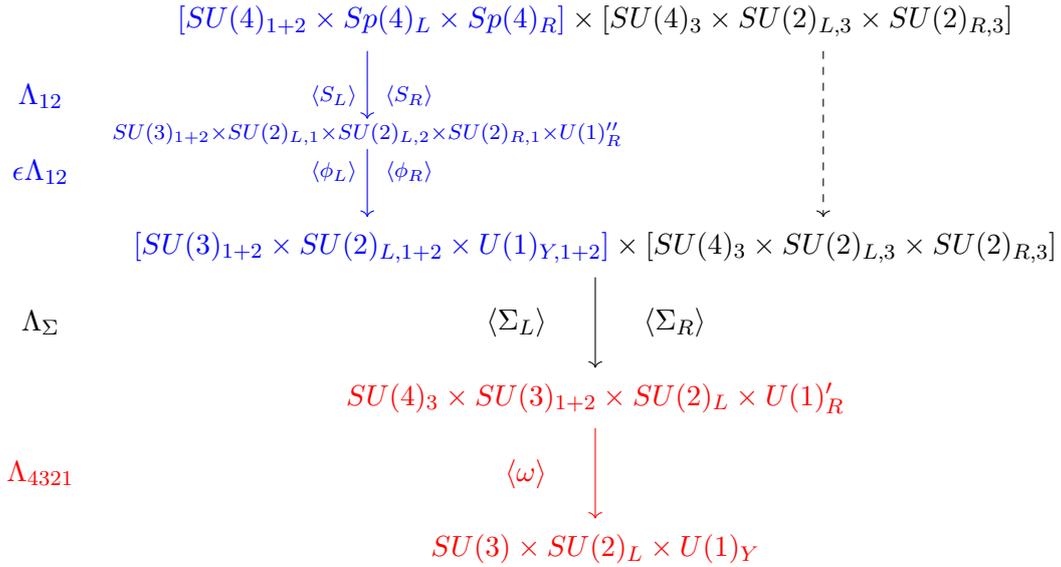

The model will explain the structure of fermion masses and mixings, while also producing third-family aligned $U_1$ leptoquarks that offer the best combined explanation of the $B$-physics anomalies, by breaking this large gauge symmetry down to the SM in a number of stages. The symmetry breaking pattern in this model is shown in Fig.~\ref{fig:SSB}. We do not attempt to explain the pattern of neutrino masses and mixings in this paper, although it is reasonable to suppose that a form of see-saw mechanism can deliver very light neutrinos. We postpone a detailed study of the neutrino sector for future work.

\section{Dynamical generation of the Yukawa sector}\label{dyngenyuk}

\subsection{Electroweak light-flavour unification} \label{sec:light-flavour}

In the 1-2 sector, we use electroweak flavour unification (EWFU) and its structured breaking to generate hierarchies in the Yukawa textures that can accommodate the data. (These symmetry breaking steps are coloured \textcolor{blue}{blue} in Fig.~\ref{fig:SSB}.) This is a simplified, 2-family version of the mechanism that was first proposed in Ref.~\cite{Davighi:2022fer} (see also~\cite{Davighi:2022vpl}).

As in Ref.~\cite{Davighi:2022fer}, the idea is to first ‘deconstruct’ the flavour-unified gauge symmetry at a high scale using a pair of scalars transforming in the representations $S_L \sim {\bf (1,5,1)}$ and $S_R \sim {\bf (\bar{4},1,4)}$ of $G_{12}$, before using scalars $\Phi_L$ and $\Phi_R$ in 2-index antisymmetrized representations (of $Sp(4)_L$ and $Sp(4)_R$ respectively) to ‘link’ the deconstructed gauge factors.
In a little more detail, the key steps in the mechanism can be summarized as follows.  

\subsubsection{Sequential symmetry breaking (part I)}

\paragraph{A. Deconstruction of electroweak symmetry, at scale $\Lambda_{12}$.} 
First, the gauge-flavour unified symmetry is `deconstructed'  around a high scale $\Lambda_{12}$, which is so labelled because it is the scale at which the structure of the light-family Yukawa couplings are generated. This is done via a pair of scalars transforming in the representations $S_L \sim {\bf (1,5,1)}$ and $S_R \sim {\bf (\bar{4},1,4)}$ of $G_{12}$. Explicitly, the breaking pattern is
\begin{align}
&\langle S_L \rangle:\,  Sp(4)_L  &&\longrightarrow  SU(2)_{L,1} \times SU(2)_{L,2} \, , \\
&\langle S_R \rangle:\, SU(4)_{1+2}\times Sp(4)_R &&\longrightarrow SU(3)_{1+2} \times SU(2)_{R,1} \times U(1)_{R}^{\prime\prime}\, ,
\end{align}
where $U(1)_R^{\prime\prime}$ acts as $B-L$ on the right-handed first family, and as hypercharge on all other light fermions. This breaking can be  achieved using vevs\footnote{The fact that the vev of $S_L\sim {\bf 5}$ deconstructs $Sp(4)_L \to SU(2)_{L,1} \times SU(2)_{L,2}$ is easily understood at the Lie algebra level by recalling that $\mathfrak{sp}(4) \cong \mathfrak{so}(5)$. The field $S_L \sim {\bf 5}$ transforms in the fundamental vector representation of $\mathfrak{so}(5)$, and so a generic vev breaks $\mathfrak{so}(5)$ to an $\mathfrak{so}(4) \cong \mathfrak{su}(2) \oplus \mathfrak{su}(2)$ subalgebra. At the group level, this translates to the breaking $Sp(4) \to SU(2)\times SU(2)$. It is straightforward to show that the vev direction (\ref{eq:SLSRvevs}) breaks $Sp(4)$ to the particular $SU(2)_{L,1}\times SU(2)_{L,2}$ subgroup that we want to preserve.} 
\be \label{eq:SLSRvevs}
\langle S_L \rangle
=\alpha_L \Lambda_{12} 
\begin{pmatrix}
0 & 0 & 1 & 0 \\
0 & 0 & 0 & -1 \\
-1 & 0 & 0 & 0 \\
0 & 1 & 0 & 0
\end{pmatrix}, 
\qquad
\langle S_R \rangle = \alpha_R \Lambda_{12} 
\begin{pmatrix}
0&0&0&0\\
0&0&0&0\\
0&0&0&0\\
0&1&0&0
\end{pmatrix}
\ee
where $\alpha_L$ and $\alpha_R$ are dimensionless numbers (that allow for the symmetry breaking in the left- and right-sectors to occur at slightly different scales, in the vicinity of $\Lambda_{12}$). 
Here, the vev of $S_L$, which transforms in an antisymmetrized 2-index representation of $Sp(4)_L$, is written as a 4-by-4 antisymmetric matrix. More precisely, the vev can be written (using the basis defined in \S \ref{sec:conventions}) as $\langle S_L \rangle=\alpha_L \Lambda_{12}  (b_1 \wedge b_3 - b_2 \wedge b_4)$, where the wedge symbol `$\wedge$' denotes antisymmetrization. 
The vev of $S_R$, which transforms in the ${\bf (\bar{4},4)}$ representation of $SU(4) \times Sp(4)_R$, is written as a 4-by-4 matrix; in index notation, we have $\langle S_R \rangle=\alpha_R \Lambda_{12}  \, a_4^\ast \otimes c_2$.

\paragraph{B. Integrate out heavy Higgs fields.} 
Under the above deconstruction step, each of the Higgs fields $\H_1\sim {\bf (1,4,4)}$ and $\H_{15} \sim {\bf (15,4,4)}$ decomposes into a set of `flavoured' Higgs fields, with components 
coupling to each pair of light families (one left-handed, and one right-handed).\footnote{This kind of model-building structure, whereby a generic set of flavoured Higgs fields are ultimately responsible for generating hierarchies in the Yukawa couplings, is reminiscent of the `scalar democracy' idea proposed in Refs.~\cite{Hill:2019ldq,Hill:2019cce}. Electroweak-flavour unification is a natural UV framework for realising scalar democracy.
}
Specifically, the $Sp(4)_L \times Sp(4)_R$ bifundamental representation ${\bf (4,4)}$ decomposes under Step 1, which breaks $SU(4)_{1+2} \times Sp(4)_L \times Sp(4)_R \to SU(3)_{1+2} \times SU(2)_{L,1} \times SU(2)_{L,2} \times SU(2)_{R,1} \times U(1)_{R}^{\prime\prime}$, as
\be \label{eq:H-decomp}
\H_a \sim {\bf (4,4)}_a \mapsto \underbrace{{\bf (2,1,2)}_0}_{\H_a^{11}} \oplus \underbrace{{\bf (1,2,2)}_0}_{\H_a^{21}}
\oplus \underbrace{{\bf (2,1,1)}_{{\frac{1}{2}}}}_{\H_a^{12+}} \oplus \underbrace{{\bf (2,1,1)}_{-\frac{1}{2}}}_{\H_a^{12-}}
\oplus \underbrace{{\bf (1,2,1)}_{\frac{1}{2}}}_{\H_a^{22+}} \oplus \underbrace{{\bf (1,2,1)}_{-\frac{1}{2}}}_{\H_a^{22-}},
\ee
where we use the notation $\H_a^{ij}$ to denote an $SU(3)_{1+2}$ singlet component coming from $\H_a$, that couples to the $i^{\text{th}}$ generation of left-handed fermion, and the $j^{\text{th}}$ generation of right-handed fermion.\footnote{The notation $\H_{15}^{ij}$ therefore denotes an $SU(3)_{1+2}$ singlet components originating from the decomposition of the adjoint Higgs $\H_{15}$.}
It is assumed that most of these flavoured Higgs components are heavy, except for the $\H_a^{22\pm}$ components which we suppose are lighter, and which will ultimately mix with the physical Higgses (see \S \ref{sec:Higgs-stability}). 
It is this presumed structure of the quadratic Higgs mass terms that {\em defines} the second generation; modulo these scalar mass terms, the theory is at this point still invariant under permuting the family labels of left-handed light fermions.

Concretely, denoting with $M^{ij}_a$ the mass of field $\H_a^{ij(\pm)}$, we assume the scalar potential is such that
\begin{align}
M^{11}_a &\approx M^{12}_a \approx M^{21}_a \approx \Lambda_{12}, \\
M^{22}_a &\text{~somewhat lighter}\, .
\end{align}
The heavy components are all integrated out at the scale $\Lambda_{12}$. This generates many EFT couplings, including higher-dimensional Yukawa operators that couple the remaining dynamical Higgs fields $\H^{22}_a$ (which will mix with the physical Higgs) to the first generation fermions. At the level of the EFT, one can already see that gauge invariance requires one insertion of $\Phi_L$ ($\Phi_R$) to couple $\H^{22}_a$ to a left-handed (right-handed) first generation fermion. Thus, schematically, the effective Yukawa couplings have the form
\be \label{eq:Y12-schem}
\cL_{\text{EFT}} \supset  
\bar{f}_{L,2} \H^{22}_a f_{R,2}
+\frac{\Phi_{L}}{\Lambda_{12}}\bar{f}_{L,1} \H^{22}_a f_{R,2}
+\frac{\Phi_{R}}{\Lambda_{12}}\bar{f}_{L,2} \H^{22}_a f_{R,1}
+\frac{\Phi_{L}\Phi_R}{\Lambda_{12}^2}\bar{f}_{L,1} \H^{22}_a f_{R,1}\, ,
\ee
for each fermion type $f\in\{u,d,e\}$. We give the precise expressions in Eqs.~(\ref{eq:Y22}--\ref{eq:Y11}).

\paragraph{C. Break to the (light-flavour) SM, at scales $\epsilon_{L,R}\, \Lambda_{12}$.} 
Finally, there are the remaining dynamical scalars $\Phi_L$ (real) and $\Phi_R$ (complex), which are in 2-index antisymmetrized representations (the $\bf{5}$) of $Sp(4)_L$ and $Sp(4)_R$ respectively that can be represented by antisymmetric 4-by-4 matrices. These scalar fields acquire vevs to `link' the deconstructed gauge factors. 
The required vevs are embedded as
\begin{align}
\langle \Phi_L \rangle = \Lambda_{12} \epsilon_L
\begin{pmatrix}
0&0&0&1\\
0&0&1&0\\
0&-1&0&0\\
-1&0&0&0
\end{pmatrix},
\qquad
\langle \Phi_R \rangle = \Lambda_{12} \epsilon_R
\begin{pmatrix}
0&0&0&z_{+}\\
0&0&z_{-}&0\\
0&-z_{-}&0&0\\
-z_{+}&0&0&0
\end{pmatrix}\, ,
\end{align}
where $z_{\pm}$ are (independent) dimensionless, order-1 $\mathbb{C}$-numbers.
In our index notation these vevs are 
\begin{align}
\langle \Phi_L \rangle &= \Lambda_{12} \epsilon_L (b_1\wedge b_4 + b_2\wedge b_3) =: \langle \phi_L \rangle, \\ 
\langle \Phi_R \rangle &= \underbrace{\Lambda_{12} \epsilon_{R} z_{+} c_1\wedge c_4}_{\langle\phi_{R+}\rangle} + \underbrace{\Lambda_{12} \epsilon_{R} z_{-} c_2\wedge c_3}_{\langle{\phi}_{R-}\rangle}\, 
\end{align}
where, in the second line, it is convenient to identify specific components of $\Phi_{L,R}$ with distinct link fields $\phi_L$, $\phi_{R+}$ and $\phi_{R-}$ in the broken phase. These transform in the representations $\phi_L \sim {\bf (2,2)}$ of $SU(2)_{L,1}\times SU(2)_{L,2}$, and $\phi_{R\pm} \sim {\bf 2}_{\pm\frac{1}{2}}$ of $SU(2)_{R,1} \times U(1)_R^{\prime\prime}$.
The factors $\epsilon_L$ and $\epsilon_{R}$ are small, $\R$-valued parameters, which parametrize the ratios of energy scales between the $\Phi_{L,R}$ condensation scale and the heavier scale $\Lambda_{12}$ at which the heavy Higgs components are integrated out. 

These vevs trigger the symmetry breaking
\begin{align}
\langle \Phi_L \rangle:\, &SU(2)_{L,1} \times SU(2)_{L,2} \longrightarrow SU(2)_{L,1+2}, \\
\langle \Phi_R \rangle:\, &SU(2)_{R,1} \times U(1)_{R}^{\prime\prime} \longrightarrow U(1)_{Y,1+2} \, ,
\end{align}
the result of which is $\text{SM}_{1+2} :=  SU(3)_{1+2}\times SU(2)_{L,1+2}\times U(1)_{Y,1+2}$, the (light-flavour) SM gauge symmetry.
Once this condensation of $\Phi_L$ and $\Phi_R$ occurs, the EFT Yukawa operators generated in Step B match onto dimension-4 Yukawa couplings of the first generation fermions to the field $\H^{22}_a$, with in-built EFT suppression factors of $\epsilon_L$ and $\epsilon_{R\pm}$.

\subsubsection{EFT matching for light Yukawas}

Having described the essential elements of the EWFU mechanism for generating Yukawa hierarchies, which we here adapt from Ref.~\cite{Davighi:2022fer} to the 2-flavour sector, we now give the details of the EFT matching.

\begin{figure}[htbp]
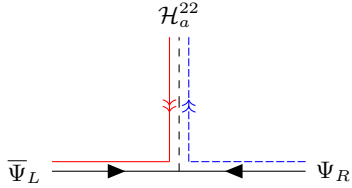

\begin{center}
\colordiagram{
    \fermions
    \HiggsII{L3}{R3}
    \LII{L0}{L3}
    \RII{R0}{R3}
}
\end{center}
\caption{Feynman diagram representing the direct coupling of second generation fermions to the $\H^{22}_a$ Higgs components, which ultimately mix with the $H_a$ fields and result in the 2-2 Yukawa couplings. \label{fig:22}}
\end{figure}

\begin{figure}[htbp]
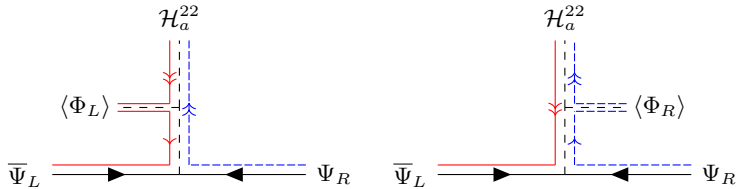

\begin{center}
\colordiagram{
    \fermions
    \HiggsII{L3}{R3}
    \phiLI{c1}
    \LI{L0}{L1}
    \LII{L2}{L3}
    \RII{R0}{R3}
}
\colordiagram{
    \fermions
    \HiggsII{L3}{R3}
	\phiRI{c1}
    \LII{L0}{L3}
    \RI{R0}{R1}
    \RII{R2}{R3}
}
\end{center}
\caption{Feynman diagrams that contribute to the 1-2 (left) and 2-1 (right) elements of the Yukawa matrices, once the heavy Higgs components running along the internal lines are integrated out at $\Lambda_{12}$. \label{fig:12}}
\end{figure}

\begin{figure}[htbp]
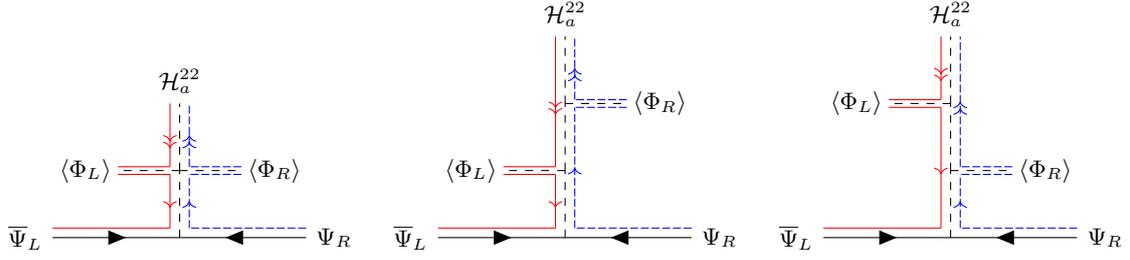

\begin{center}
\colordiagram{
    \fermions
    \HiggsII{L3}{R3}
    \phiLI{c1}
	\phiRI{c1}
    \LI{L0}{L1}
    \LII{L2}{L3}
    \RI{R0}{R1}
    \RII{R2}{R3}
}
\colordiagram{
    \fermions
    \HiggsIII{L4}{R4}
    \phiLI{c1}
	\phiRI{c2}
    \LI{L0}{L1}
    \LII{L2}{L4}
    \RI{R0}{R1}
    \RII{R2}{R4}
}
\colordiagram{
    \fermions
    \HiggsIII{L4}{R4}
    \phiLI{c2}
	\phiRI{c1}
    \LI{L0}{L1}
    \LII{L2}{L4}
    \RI{R0}{R1}
    \RII{R2}{R4}
}
\end{center}
\caption{Feynman diagrams that contribute to the 1-1 element of the Yukawa matrices, once the heavy Higgs components running along the internal lines are integrated out at $\Lambda_{12}$. The presence of the diagram on the left is essential in order to provide enough freedom to parametrize the 1-1 entry independently of the 1-2 and 2-1 entries.\label{fig:11}}
\end{figure}

The effective Yukawa operators that we wrote schematically in Eq.~(\ref{eq:Y12-schem}) are automatically generated upon integrating out the heavy components of $\H^{ij}_a$ at scale $\Lambda_{12}$, provided the UV model contains the following gauge invariant cubic and quartic scalar interactions 
\begin{align}
V(\Phi,H) \supset &\sum_{a\in\{1,15\}} 
\Lambda_{12}\left[ \beta_L^a \Tr ( \H_a^{\ast} \Phi_L  \H_a)
+  \beta_R^a \Tr ( \H_a^{\ast}  \H_a  \Phi_R) 
+ \beta_R^{a\ast} \Tr ( \H_a^{\ast}   \H_a \Phi_R^\ast)\right] \nonumber \\
+ &\sum_{a\in\{1,15\}} \beta_{LR}^a \Tr ( \H_a^{\ast} \Phi_L\H_a  \Phi_R) 
+\beta_{LR}^{a\ast} \Tr (\H_a^{\ast} \Phi_L  \H_a \Phi_R^\ast)\, ,
\end{align}
which couple the $\H_a$ Higgs fields (and their conjugates) to the symmetry breaking scalars $\Phi_L$ and $\Phi_R$.
Given these scalar interactions, one can draw the Feynman diagrams shown in Figs.~\ref{fig:22}--\ref{fig:11}; upon integrating out the $\mathcal{H}_a^{ij}$ components, which run as internal propagators, one gets the desired higher-dimension Yukawa operators.

If we decompose the fermion fields into their family-deconstructed components, {\em i.e.} those states relevant after all the symmetry breaking steps detailed in \S \ref{sec:light-flavour}, the resulting effective Yukawa operators obtained by EFT matching can be organised by increasing mass dimension, as follows.
\medskip
\\
\noindent 
\underline{Dimension 4 (see Fig.~\ref{fig:22}):}
\begin{align} 
\cL_{\text{EFT}} \supset \quad
&\overline{Q}_{L,2} \left(y_1^l \H^{22-}_1 + \overline{y}_1^l \H^{22+\ast}_1 + y_{15}^l \H^{22-}_{15} + \overline{y}_{15}^l \H^{22+\ast}_{15} \right)  u_{R,2}\,  +\text{h.c.}  \no\\
+ &\overline{Q}_{L,2} \left(y_1^l \H^{22+}_1 + \overline{y}_1^l \H^{22-\ast}_1 + y_{15}^l \H^{22+}_{15} + \overline{y}_{15}^l \H^{22-\ast}_{15} \right)  d_{R,2}\,  +\text{h.c.}  \no \\
+ &\overline{L}_{L,2} \left(y_1^l \H^{22+}_1 + \overline{y}_1^l \H^{22-\ast}_1 -3 y_{15}^l \H^{22+}_{15} -3 \overline{y}_{15}^l \H^{22-\ast}_{15} \right) e_{R,2}\, +\text{h.c.} \, . \label{eq:Y22}
\end{align}
To simplify the notation, it is convenient to define :
\be
\begin{pmatrix} \H_a^u \\ \overline{\H}_a^u \end{pmatrix} = \mathcal{R}_a \begin{pmatrix} \H_a^{22-} \\ H_a^{22+\ast} \end{pmatrix}\, ,
\qquad 
\begin{pmatrix} \H_a^d \\ \overline{\H}_a^d \end{pmatrix} = \mathcal{R}_a \begin{pmatrix} \H_a^{22+} \\ \H_a^{22-\ast} \end{pmatrix}\,,
\ee
where the unitary matrix $\mathcal{R}_a$ is 
\be
\mathcal{R}_a=\frac{1}{\sqrt{|y_a^l|^2+ |\bar{y}_a^l |^2}} \begin{pmatrix} y_a^l & \bar{y}_a^l \\ -\left(\bar{y}_a^l\right)^\ast & \left( y_a^l\right)^\ast \end{pmatrix} \, .
\ee
This way only the combinations $\H_a^{u,d}$ appear in the Yukawa interaction  (\ref{eq:Y22}).
The latter assumes the simple form
\be
\cL_{\text{EFT}} \supset \sum_{a\in\{1,15\}} \tilde{y}_a^l \left( 
\overline{Q}_{L,2} \H^{u}_{a} u_{R,2} + 
\overline{Q}_{L,2} \H^{d}_{a} d_{R,2} + 
\overline{L}_{L,2} \Gamma_a \H^{d}_{a} e_{R,2}
\right) +\text{~h.c.}\,,
\ee
where  $\Gamma_1 \equiv 1$ and $\Gamma_{15} \equiv -3$ encode the different treatment of the $SU(3)$-singlet piece ({\em i.e.} the leptons) contained in $SU(4)$,
and 
\be
\tilde{y}_a^l = \sqrt{|y_a^l|^2+ |\bar{y}_a^l |^2}\, .
\ee
Proceeding in a similar manner, the subleading contributions to the effective Yukawa interaction generated by high-dimensional 
operators are as follows.\\
\vskip 1 pt
\noindent 
\underline{Dimension 5 (see Fig.~\ref{fig:12}):}
\begin{align} \label{eq:Y12}
\cL_{\text{EFT}} \supset \sum_{a\in\{1,15\}}
\tilde{y}_a^l \Big[ &\overline{u}_{L,1} \frac{\beta_L^a \phi_L}{\Lambda_{12}}\H^u_a\,  u_{R,2}\, 
+\overline{u}_{L,2} \left(\frac{\beta_R^a \phi_{R+}}{\Lambda_{12}} + \frac{\beta_R^{a\ast} \phi_{R-}^\ast}{\Lambda_{12}} \right)\H^u_a\,  u_{R,1} \nonumber \\
+&\overline{d}_{L,1} \frac{\beta_L^a \phi_L}{\Lambda_{12}}\H^d_a\, d_{R,2} + \overline{d}_{L,2} \left(\frac{\beta_R^a \phi_{R+}^\ast}{\Lambda_{12}} + \frac{\beta_R^{a\ast} \phi_{R-}}{\Lambda_{12}} \right)\H^d_a\, d_{R,1} \nonumber \\
+& (d \leftrightarrow e\,, \tilde{y}_a^l \leftrightarrow \Gamma_a \tilde{y}_a^l) \Big] +\text{~h.c.}\, .
\end{align}
 \underline{Dimension 6 (see Fig.~\ref{fig:11}):}
\begin{align} \label{eq:Y11}
\cL_{\text{EFT}} \supset \sum_{a\in\{1,15\}}\tilde{y}_a^l \Big[&
\overline{u}_{L,1} \left( \frac{(\beta_{LR}^a+2\beta_L^a\beta_R^a) \phi_L\phi_{R+}}{\Lambda_{12}^2}+ \frac{(\beta_{LR}^{a\ast}+2\beta_L^a\beta_R^{a\ast})  \phi_L\phi_{R-}^\ast}{\Lambda_{12}^2}\right)\H^u_a\, u_{R,1} \nonumber \\
+&\overline{d}_{L,1} \left( \frac{(\beta_{LR}^a+2\beta_L^a\beta_R^a) \phi_L\phi_{R+}^\ast}{\Lambda_{12}^2}+ \frac{(\beta_{LR}^{a\ast}+2\beta_L^a\beta_R^{a\ast})  \phi_L\phi_{R-}}{\Lambda_{12}^2}\right) 
\H^d_a\, d_{R,1}\nonumber \\
+&(d \leftrightarrow e\,, \tilde{y}_a^l \leftrightarrow \Gamma_a \tilde{y}_a^l) \Big] +\text{~h.c.} \, .
\end{align}

\bigskip
To summarize the 1-2 sector generated by this mechanism, for all three types of SM fermion $f \in \{u,d,e\}$, the Yukawa couplings $Y^{22}_f$ between the light families and the $\H^{22}$ Higgs fields have the hierarchical structure
\begin{equation}
Y^{22}_f \sim 
\begin{pmatrix}
\epsilon_L \epsilon_R & \epsilon_L \\
\epsilon_R & 1
\end{pmatrix}\, .
\end{equation}
This is the 2-flavour version of the EWFU structure derived in~\cite{Davighi:2022fer}.

\subsection{Mixing with the third family}

At this point the construction of the Yukawa sector departs from the EWFU mechanism of~\cite{Davighi:2022fer} (which essentially replicates the structure we have just described in all three families). For us, the third family is treated differently in the UV, coupling to its own decoupled Pati--Salam gauge factor $G_3$, as in Eqs. (\ref{eq:G}--\ref{eq:G3}).

Before we explain how the remaining Yukawa couplings (that mix with the third family) are generated, we must explain the final symmetry breaking steps needed to take us from $\text{SM}_{1+2} \times G_3$ down to $\text{SM}_{1+2+3}$, the SM gauge symmetry. Continuing our labelling from that of \S \ref{sec:light-flavour}, these steps are the following.

\subsubsection{Sequential symmetry breaking (part II)}

\paragraph{D. Linking of electroweak symmetry, at scale $\Lambda_{\Sigma}$.} 

The lowest breaking step (E) will, by design, mimic the symmetry breaking in so-called `4321 models'~\cite{DiLuzio:2017vat,Greljo:2018tuh,DiLuzio:2018zxy}. To make contact with this requires we first link together the electroweak symmetries, which remain partially deconstructed. This step resembles a similar linking step in the `Pati--Salam cubed' UV completion~\cite{Bordone:2017bld,Bordone:2018nbg,Fuentes-Martin:2020pww} of 4321.
This breaking step is triggered by two complex scalar fields, which start life in representations $\Sigma_L \sim \bf{(1,4,1)} \otimes \bf{(1,2,1)}$ and $\Sigma_R \sim \bf{(1,1,4)} \otimes \bf{(1,1,2)}$ of the UV gauge symmetry $G$ (\ref{eq:G}).

Representing both these bifundamentals as 2-by-4 matrices (for which each row transforms as an $Sp(4)_{L/R}$ fundamental), appropriate vevs are
\be \label{eq:vevSigmas}
\langle \Sigma_L \rangle = \gamma_L\Lambda_\Sigma {\bf M}, \qquad
\langle \Sigma_R \rangle = \gamma_R\Lambda_\Sigma {\bf M},
\ee
where the matrix ${\bf M}$ is 
\be \label{eq:vevSigmas2}
{\bf M} = \begin{pmatrix}
0&1&0&0\\
0&0&0&1
\end{pmatrix}\, .
\ee
In our index notation, this is equivalent to $\langle \Sigma_L \rangle = \gamma_L\Lambda_\Sigma (b_2\otimes B_1 + b_4 \otimes B_2)$ and $\langle \Sigma_R \rangle = \gamma_R\Lambda_\Sigma (c_2\otimes C_1 + c_4 \otimes C_2)$. The result is to break
\be
SU(2)_{L,1+2} \times U(1)_{Y,1+2} \times SU(2)_{L,3} \times SU(2)_{R,3} \longrightarrow SU(2)_L \times U(1)_{R}^\prime,
\ee
leaving the electroweak factor of the 4321 model. Here $U(1)_R^\prime$ acts as hypercharge on the light families, and as $U(1)_R$ ({\em i.e.} the subgroup of $SU(2)_R$ generated by $\sigma_3$) on the third family.

\definecolor{Gray}{gray}{0.9}
\begin{table}
\begin{center}{\small
\begin{tabular}{|c|c|cccc|}
\hline
 & Field ($i,j=1,2$) & $SU(4)_3$ & $SU(3)_{1+2}$ & $SU(2)_L$ & $U(1)^\prime_R$ \\
\hline
SM Fermions (chiral) & $Q_L^{i}$ & $\bf{1}$ & $\bf{3}$ & $\bf{2}$ & $1/6$  \\
										  & $u_R^{i}$ & $\bf{1}$ & $\bf{3}$ & $\bf{1}$ & $2/3$  \\
										  & $d_R^{i}$ & $\bf{1}$ & $\bf{3}$ & $\bf{1}$ & $-1/3$  \\
										  & $L_L^{i}$ & $\bf{1}$ & $\bf{3}$ & $\bf{2}$ & $-1/2$  \\
										  & $e_R^{i}$ & $\bf{1}$ & $\bf{3}$ & $\bf{1}$ & $-1$  \\
										  & $\nu_R^{i}$ & $\bf{1}$ & $\bf{3}$ & $\bf{1}$ & $0$  \\
										  & $\Psi_L^{3}$ & $\bf{4}$ & $\bf{1}$ & $\bf{2}$ & $0$  \\
										  & $\Psi_{R,u}^{3}$ & $\bf{4}$ & $\bf{1}$ & $\bf{1}$ & $1/2$  \\
										  & $\Psi_{R,d}^{3}$ & $\bf{4}$ & $\bf{1}$ & $\bf{1}$ & $-1/2$  \\
\hline
\hline
Vector-like fermion & $\xi_{u}^{i}$ & $\bf{1}$ & $\bf{3}$ & $\bf{1}$ & $2/3$ \\
                          & $\xi_{d}^{i}$ & $\bf{1}$ & $\bf{3}$ & $\bf{1}$ & $-1/3$ \\
                          & $\xi_{e}^{i}$ & $\bf{1}$ & $\bf{1}$ & $\bf{1}$ & $-1$ \\
                          & $\xi_{\nu}^{i}$ & $\bf{1}$ & $\bf{1}$ & $\bf{1}$ & $0$ \\
\hline
\hline
		   Higgses (light only) & $\mathcal{H}_a^{22+}$ & $\bf{1}$ & $\bf{1}$ & $\bf{2}$ & $+1/2$ \\
                    & $\mathcal{H}_a^{22-}$ & $\bf{1}$ & $\bf{1}$ & $\bf{2}$ & $-1/2$ \\
                    & $R_2$ & $\bf{1}$ & $\bf{3}$ & $\bf{2}$ & $7/6$ \\
                    & $\tilde{R}_2$ & $\bf{1}$ & $\bf{3}$ & $\bf{2}$ & $1/6$ \\
	\rowcolor{Gray}
                    & $H_{1}^+$ & $\bf{1}$ & $\bf{1}$ & $\bf{2}$ & $1/2$ \\
	\rowcolor{Gray}
                    & $H_{15}^+$ & $\bf{15}$ & $\bf{1}$ & $\bf{2}$ & $1/2$ \\
	\rowcolor{Gray}
                    & $H_{1}^-$ & $\bf{1}$ & $\bf{1}$ & $\bf{2}$ & $-1/2$ \\
	\rowcolor{Gray}
                    & $H_{15}^-$ & $\bf{15}$ & $\bf{1}$ & $\bf{2}$ & $-1/2$ \\
\hline
\hline
Symmetry breaking scalars & $\omega_{uu}^{i3}$ & $\bar{\bf{4}}$ & $\bf{3}$ & $\bf{1}$ & $7/6$ \\
	\rowcolor{Gray}
										& $\omega_{ud}^{i3}\,, \  \omega_{ud}^{3i}$  & $\bar{\bf{4}}$ & $\bf{3}$ & $\bf{1}$ & $1/6$ \\
										& $\omega_{dd}^{i3}$ & $\bar{\bf{4}}$ & $\bf{3}$ & $\bf{1}$ & $-5/6$ \\
										& $\omega_{\nu\nu}^{i3}$ & $\bar{\bf{4}}$ & $\bf{1}$ & $\bf{1}$ & $1/2$ \\
	\rowcolor{Gray}
										& $\omega_{e\nu}^{i3}\,, \  \omega_{\nu e}^{3i}$ & $\bar{\bf{4}}$ & $\bf{1}$ & $\bf{1}$ & $-1/2$ \\
										& $\omega_{ee}^{i3}$ & $\bar{\bf{4}}$ & $\bf{1}$ & $\bf{1}$ & $-3/2$ \\
\hline
\end{tabular}
}
\end{center}
\caption{Field content of the model before 4321 breaking. Of the scalar fields (listed in the last two blocks), those that acquire non-vanishing vevs are indicated by grey shading.
} \label{tab:Matter_Content_4321}
\end{table}

As for Step C above, we can decompose these symmetry breaking scalars into components of the intermediate gauge symmetry that remains dynamical at this scale, namely $\text{SM}_{1+2} \times G_3$. The field $\Sigma_L$ just splits into a pair of bidoublets,
\be \label{eq:SigmaL_decomp}
\Sigma_L \longrightarrow {\bf (2,2)}^{\oplus 2} \quad \text{of} \quad SU(2)_{L,1+2} \times SU(2)_{L,3}\, ,
\ee
while
\be   \label{eq:SigmaR_decomp}
\Sigma_R \longrightarrow (+1/2, {\bf 2})^{\oplus 2} \oplus (-1/2, {\bf 2})^{\oplus 2} \quad \text{of} \quad U(1)_{Y,1+2} \times SU(2)_{R,3} \, .
\ee
The fact that all the fields in these decompositions are duplicated is because of their origin in the gauge-flavour unified $Sp(4)_{L(R)}$ symmetries; there is {\em one copy} of every field {\em for each light family}. The vevs (\ref{eq:vevSigmas}--\ref{eq:vevSigmas2}) sit in $Sp(4)$ components corresponding to the second family; while other choices would have achieved the same symmetry breaking pattern (which is obvious because the decompositions (\ref{eq:SigmaL_decomp}--\ref{eq:SigmaR_decomp}) carry no trace of the light family index, which has now been linked together in $\text{SM}_{1+2}$),
we will see that the vevs of $\Sigma_{L,R}$ play another role. This second role is in the mixing of the different Higgs fields to produce the physical mass eigenstates; as we alluded to above, this is where the second and first family are distinguished, and so the choice (\ref{eq:vevSigmas2}) is important. See \S \ref{sec:SigmaMixHiggs}.

\paragraph{E. Breaking 4321 to the Standard Model, at scale $\Lambda_{4321}$.}

At this point, the gauge symmetry is that of the 4321 model, namely
\be \label{eq:4321}
SU(4)_3 \times SU(3)_{1+2} \times SU(2)_L \times U(1)_R^\prime\, .
\ee
Even though the gauge group (and its action on the SM chiral fermions) is the same as for established 4321 models~\cite{DiLuzio:2017vat,DiLuzio:2018zxy,Greljo:2018tuh}, this model features a different scalar sector and choice of vector-like fermion, which we record in Table~\ref{tab:Matter_Content_4321}.

The remaining symmetry breaking step is to the SM. The final scalar field $\omega$, whose vev triggers this breaking, transforms in the representation $\omega \sim {\bf (4,1,4)} \otimes {\bf (\bar{4},1,2)}$ of the UV gauge symmetry $G$ (\ref{eq:G}).
At this point it  is not so enlightening to record how the vev is embedded in the UV field, so we begin by decomposing this field under the 4321 group (\ref{eq:4321}). We have\footnote{Note that the fields here labelled $\widehat{\omega}_{3(1)}$ themselves denote {\em reducible} representations of 4321, which we use as a notational convenience to gather together the $SU(3)_{1+2}$ triplets (singlets). }
\begin{align} \label{eq:Omh}
\omega &\rightarrow \widehat{\omega}_3 \oplus \widehat{\omega}_1\, , \\
\widehat{\omega}_3 &\sim \{ \underbrace{{\bf (\bar{4},3,1},7/6)}_{\omega_{uu}^{\textcolor{red}{i}3}} \oplus  {\underbrace{{\bf (\bar{4},3,1},1/6)}_{\omega_{ud}^{\textcolor{red}{i}3},\,\, \omega_{ud}^{3\textcolor{red}{i}}}}^{\oplus 2} \oplus \underbrace{{\bf (\bar{4},3,1},-5/6)}_{\omega_{dd}^{\textcolor{red}{i}3}}   \}^{\textcolor{red}{\oplus 2}}\, , \nonumber \\
\widehat{\omega}_1 &\sim \{ \underbrace{{\bf (\bar{4},1,1},1/2)}_{\omega_{\nu\nu}^{\textcolor{red}{i} 3}} \oplus  {\underbrace{{\bf (\bar{4},1,1},-1/2)}_{\omega_{\nu e}^{\textcolor{red}{i}3},\,\, \omega_{\nu e}^{3\textcolor{red}{i}}}}^{\oplus 2} \oplus \underbrace{{\bf (\bar{4},1,1},-3/2)}_{\omega_{ee}^{\textcolor{red}{i}3}}   \}^{\textcolor{red}{\oplus 2}}\, . \nonumber
\end{align}
Our notation here needs a little explanation. The subscript label indicates the pair of fermion types to which that $\omega$ component couples; for example, the fields $\omega_{uu}^{i3}$ are colour triplets and so couple to di-quark pairs, and their $U(1)_R^\prime$ charge of $+7/6$ means they couple specifically to pairs of up-type quarks. The superscript label then indicates the family indices of the fermion pair to which that $\omega$ component couples. One of these is always a third family label, and the other is always a light family label which can run over $\textcolor{red}{i \in \{1,2\}}$; this is because the UV field $\omega$ is in a bifundamental representation of $Sp(4)_{R,12} \times SU(2)_{R,3}$, and the duplication labelled by $i$ comes from the fact that $Sp(4)_{R,12}$ stores a light family index. For the fields with `mixed couplings', {\em i.e.} to one up-type and one down-type quark, there is a further duplication due to the fact that there are components coupling to $\{$second family down-type, third family up-type$\}$, and vice-versa. Thus, for example, there are four of these scalars with quantum numbers ${\bf (\bar{4},3,1},1/6)$. 
To emphasize this, we will at times use notation 
\beq \label{eq:om_combs}
\omega_{ud}^{13} \equiv \omega^{ub}\, , \qquad 
\omega_{ud}^{23} \equiv \omega^{cb}\, , \qquad 
\omega_{ud}^{31} \equiv \omega^{td}\, , \qquad 
\omega_{ud}^{32} \equiv \omega^{ts}\, ,
\eeq
labelling the quark (or lepton) flavours explicitly.

The 4321 gauge symmetry is broken down to the SM gauge symmetry, {\em i.e.}
\be
SU(4)_3 \times SU(3)_{1+2} \times SU(2)_L \times U(1)_R^\prime \longrightarrow SU(3) \times SU(2)_L \times U(1)_Y\, ,
\ee
if any of the components $\omega_{ud}^{i3,\, 3i}$ (and possibly $\omega_{\nu e}^{i3,\, 3i}$) acquire non-zero vevs in the directions
\be \label{eq:vevomega}
\langle \omega_{ud}^{i3,\, 3i} \rangle = \frac{v_3}{\sqrt{2}} \begin{pmatrix}
1&0&0\\
0&1&0\\
0&0&1\\
0&0&0
\end{pmatrix}\,, \qquad
\langle \omega_{\nu e}^{i3,\, 3i} \rangle = \frac{v_1}{\sqrt{2}} \begin{pmatrix}
0\\
0\\
0\\
1
\end{pmatrix}\,,
\ee
mimicking the original 4321 setup of~\cite{DiLuzio:2017vat}, where we represent $\omega_{ud}^{i3,\, 3i}$ and $\omega_{\nu e}^{i3,\, 3i}$ as a 4-by-3 matrix and a 4-vector respectively. 

However, as a point of departure from other 4321 models in the literature, we emphasize that in this model there are four copies of each of these scalars, as listed in (\ref{eq:om_combs}). Which of these copies acquire symmetry-breaking vevs in fact has physical consequences; 
after integrating out the vector-like fermion (VLF) $\Xi$ (see the following Section \ref{sect:Yuk_mix}), the choice of vev-acquiring-$\omega$ fields determines which 2-3 Yukawa elements are populated in our EFT (up to subleading corrections due to mixing effects between the scalars). While all choices are equally natural, phenomenological reasons suggest we take the components
\be \label{eq:omega_choice}
\omega_{ud}^{23} \equiv \omega^{cb} \qquad \text{and} \qquad \omega_{\nu e}^{32} \equiv \omega^{\nu_\tau \mu}
\ee
as the (only) ones that get the vevs indicated in Eq.~(\ref{eq:vevomega}).

\subsubsection{EFT matching for third family Yukawas and mixings }
\label{sect:Yuk_mix}

Another ingredient is required to generate mixing between the light families and the third, and this is the vector-like fermion (VLF) $\Xi \sim {\bf (4,1,4)} \otimes {\bf (1,1,1)}$,
which has the same quantum numbers as the field $\Psi_R^l$. 

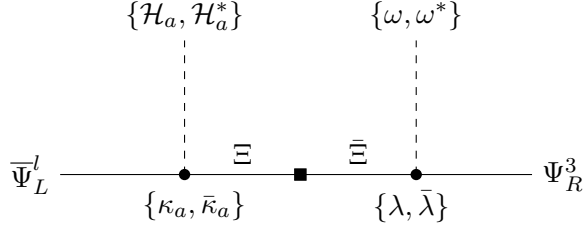
\begin{figure*}[h]
\begin{center}
\begin{tikzpicture}
\begin{feynman}
\vertex (a) { $\overline{\Psi}^l_L$};
\vertex [right=0.8in of a] (b);
\vertex [right=0.6in of b] (c);
\vertex [right=0.6in of c] (d);
\vertex [right=0.6in of d] (e) { $\Psi_R^3$};
\node at (b) [circle,fill,inner sep=1.5pt,label=below:{ $\,\,\,\, \{\kappa_a,\bar{\kappa}_a\}$}]{};
\node at (d) [circle,fill,inner sep=1.5pt,label=below:{ $\{\lambda, \bar\lambda\}\,\,$}]{};
\node at (c) [square dot,fill,inner sep=1.0pt]{};
\vertex [above=0.7in of b] (f) { $\{\H_a, \H_a^\ast\}$};
\vertex [above=0.7in of d] (g) { $\{\omega,\omega^\ast\}$};
\diagram* {
(a) -- [plain] (b)  -- [plain,edge label={ $\Xi$}] (c) -- [plain,edge label={ $\bar{\Xi}$}] (d) -- [plain] (e),
(b) -- [scalar] (f),
(d) -- [scalar] (g),
};
\end{feynman}
\end{tikzpicture}
\end{center}
\caption{Feynman diagrams, depicted in the unbroken $G_{12}\times G_3$ phase, that lead to mixing between the third family and light family fermions.
Integrating out the VLF $\Xi$ and expanding the scalar field $\omega$ about its vev gives the relevant effective Yukawa couplings mixing the light families with the third. 
The notation `$\{\kappa, \bar\kappa\}$' denotes, for example, that there are two possible couplings that can feature, connecting to either $\H_a$ or its conjugate $\H^\ast$. Recall from \S \ref{sec:dim4yuks} that the bar denotes an independent coupling.
 \label{fig:VLF} }    
\end{figure*}

This VLF permits the fundamental Yukawa interactions written above in Eq.~(\ref{eq:Xi_yukawa}), which {\em link} third family fermion fields to the light fermions via either $\mathcal{H}_a^{(\ast)}$ or $\omega$. Using these interactions, one can write down the Feynman diagram in Fig.~\ref{fig:VLF}, that links $\Psi_L^l$ with $\Psi_R^3$. Note that, given the quantum numbers of the fields in our model (in particular, given the quantum numbers of $\Xi$), there is no corresponding tree-level diagram by which we can link $\Psi_L^3$ with $\Psi_R^l$.\footnote{Such terms, which would populate the third row of the Yukawa matrices, would be generated if we further added a VLF charged under $Sp(4)_L$, which we choose not to.} 

Since the fields $\H_a$ will mix with the physical Higgs (through the components $\H_a^{22 \pm}$ introduced in (\ref{eq:H-decomp})), the Feynman diagram in Fig.~\ref{fig:VLF} will result in Yukawa couplings linking light left-handed fermions with the third family right-handed fields, populating the third column (but not the third row) of the Yukawa matrices. These terms will moreover be suppressed (with respect to the 33 entries of the Yukawa matrices) by the same overall factor by which the second family masses are suppressed, helping to explain why the mixing angles with the third family are so small.\footnote{If we had instead included a VLF charged under $G_3$, the 1-3 and 2-3 entries of the Yukawa matrices would be populated by direct couplings to the $H_1$ and $H_{15}$ Higgs fields, which recall are those Higgses charged under $G_3$. This is a less desirable option, because the smallness of the 1-3 and 2-3 quark mixing angles would have to be explained purely as a result of the heaviness of the VLF with respect to $v_{1,3}$.
}

Under 4321 the VLF decomposes as
\begin{align}
\Xi &\rightarrow \widehat{\xi}_3 \oplus \widehat{\xi}_1\, , \\
\widehat{\xi}_3 &\sim \{ \underbrace{{\bf (1,3,1},2/3)}_{\xi_u^{\textcolor{red}{i}}} \oplus \underbrace{{\bf (1,3,1},-1/3)}_{\xi_d^{\textcolor{red}{i}}}   \}^{\textcolor{red}{\oplus 2}}\, , \nonumber\\
\widehat{\xi}_1 &\sim \{ \underbrace{{\bf (1,1,1},0)}_{\xi_\nu^{\textcolor{red}{i}}} \oplus \underbrace{{\bf (1,1,1},-1)}_{\xi_e^{\textcolor{red}{i}}}   \}^{\textcolor{red}{\oplus 2}}\, , \nonumber
\end{align}
where we employ a similar notation to that used for $\omega$ in (\ref{eq:Omh}). Again, there are two copies of each 4321 representation; one set of VLFs couples to the first family fermions, and the other set to the second family, and this is labelled by the index $i$, coloured \textcolor{red}{red}. Both sets of interactions are present in the UV couplings (\ref{eq:Xi_yukawa}).

In the 4321 phase, the diagram in Fig.~\ref{fig:VLF} decomposes into separate contributions (with different mediator fields) for quarks and leptons. The $SU(3)_{1+2}$ triplet components $(\omega_{ud}, \xi_{u,d}^2)$ contribute to the quark Yukawas, while the $SU(3)_{1+2}$ singlets 
$(\omega_{\nu e}, \xi_{\nu,e}^2)$ contribute to the lepton Yukawas.

Similarly to the case of the light Yukawa couplings, it is convenient to introduce a new Higgs basis
\be
\begin{pmatrix} \left(\H_\kappa\right)_a^u \\ \left(\overline{\H_\kappa}\right)_a^u \end{pmatrix} = \mathcal{K}_a \begin{pmatrix} \H_a^{22-} \\ \H_a^{22+\ast} \end{pmatrix}\, , \qquad
\begin{pmatrix} \left(\H_\kappa\right)_a^d \\ \left(\overline{\H_\kappa}\right)_a^d \end{pmatrix} := \mathcal{K}_a \begin{pmatrix} \H_a^{22+} \\ \H_a^{22-\ast} \end{pmatrix}\, ,
\ee
where 
\be
\mathcal{K}_a:=\frac{1}{\sqrt{|\kappa_a|^2+ |\bar{\kappa}_a|^2}} \begin{pmatrix} \kappa_a & \bar{\kappa}_a \\ -\left(\bar{\kappa}_a\right)^\ast & \left( \kappa_a\right)^\ast \end{pmatrix} \,.
\ee
Defining in addition
\be
\tilde{\kappa}_a= \sqrt{|\kappa_a|^2+ |\bar{\kappa}_a |^2}\, ,
\ee
the effective heavy-light Yukawa interactions assume the form
\begin{align}
\mathcal{L} \supset \sum_{a\in\{1,15\}} &\frac{\lambda v_3 \tilde{\kappa}_a}{m_{\xi_u^2}}  \left[\overline{u}_{L,2} (\H_\kappa)_a^u \, u_{R,3} + \beta_L^a \epsilon_L\,  \overline{u}_{L,1} (\H_\kappa)_a^u \, u_{R,3}  \right] \\
+ &\frac{\lambda v_1 \tilde{\kappa}_a}{m_{\xi_e^2}}  \left[\overline{e}_{L,2} (\H_\kappa)_a^d \, e_{R,3} +  \beta_L^a \epsilon_L\, \overline{e}_{L,1} (\H_\kappa)_a^d \, e_{R,3}  \right] +\text{~h.c.} \, .
\end{align}
We remark that there is no equivalent mixing term generated in the down-quark sector, at least to this order, because of the 4321-breaking vev appearing in the $\omega^{cb}$ component, as in Eq.~(\ref{eq:omega_choice}).
It turns out that the vertex with coupling $\bar\lambda$, defined in the UV Lagrangian (\ref{eq:Xi_yukawa}), does not contribute to any non-vanishing Yukawa operator of this kind once $\omega$ is expanded about its 4321-breaking vev. \\
\indent 
The scalar potential of the model will lead the  four $SU(2)_{L,1+2}$ Higgs doublets in 
$\H^{22}_a$ (namely  $\H^{22+}_1, \H^{22-}_1, \H^{22+}_{15}, \H^{22-}_{15}$)
to have a small mixing with the corresponding $SU(2)_{L,3}$ 
Higgs doublets in $H_1$, which are the fields driving the SM electroweak symmetry breaking
(see Sect.~\ref{sec:Higgs-stability}).
As a consequence, the four neutral Higgs doublets in  $\H^{22}_a$ 
acquire a non-vanishing vev that, in general, we can write as 
\be
 \langle \H_a^{22+}  \rangle  =     \begin{pmatrix} 0 \\ \epsilon_h \eta^+_a  v  \end{pmatrix}\, ,\quad \langle \H_a^{22-}  \rangle  =     \begin{pmatrix}  \epsilon_h \eta^-_a  v\\ 0 \end{pmatrix}  \, ,
\label{eq:epsh}
\ee
where $v \approx 246$~GeV is the SM electroweak scale, $\epsilon_h \ll 1$ and $\eta^\pm_a  = \mathcal{O}(1)$.
The mass matrices of  the SM fermions then arise from the combination of couplings of the third family to $H_a$, 
yielding 33 elements of $\mathcal{O}(v)$, and couplings of all other fermion bilinears to $\H^{22}_a$, 
yielding parametrically suppressed contributions of $\mathcal{O}(\epsilon_h  v)$.

\subsection{Fermion mass and mixing angle observables}

Putting everything together from the previous Subsections, one can derive the mass matrices of the SM fermions after electroweak symmetry breaking. For each of $f \in \{u,d,e\}$, the mass matrix assumes the following hierarchical structure
\be
M_f \sim  \frac{v}{\sqrt{2} }
\begin{pmatrix}
\epsilon_h \epsilon_L \epsilon_R & \epsilon_h \epsilon_L & \epsilon_h \epsilon_L \delta_f \\
\epsilon_h \epsilon_R & \epsilon_h & \epsilon_h \delta_f \\
0 & 0 & 1
\end{pmatrix},\,
\ee
where we have introduced the quantities
\be \label{eq:deltas}
\delta_u = \frac{\lambda v_3}{m_{\xi_u^2}}\,, \qquad \delta_d \ll \delta_u 
\,,  \qquad \delta_e = \frac{\lambda v_1}{m_{\xi_e^2}}\,,
\ee
where the hierarchy between $\delta_d$ and $\delta_u$ follows from the (natural) choice in Eq.~(\ref{eq:omega_choice}), that $\omega^{cb}$ gets the vev (a comparable $\delta_d \sim \delta_u$ would be achieved if a vev was also developed by the $\omega^{ts}$ component).
Using matrix perturbation theory, one can extract the following overall scaling of the eigenvalues of such matrices, {\em i.e.} of the fermion masses in our model:
\begin{align}
m_1 &\sim \mathcal{O}(\epsilon_L \epsilon_R\, \epsilon_h)\, , \\
m_2 &\sim \mathcal{O}(\epsilon_h)\, , \\
m_3 &\sim \mathcal{O}(1)\, .
\end{align}
The CKM angles scale as
\begin{align}
V_{us} &\sim \mathcal{O}(\epsilon_L)\,, \\
V_{cb} &\sim \mathcal{O}(\epsilon_h \delta_{u})\, , \\
V_{ub} &\sim \mathcal{O}(\epsilon_L \epsilon_h \delta_{u})\, ,
\end{align}
where we emphasize that each of these CKM elements admits a complex phase. 
The precise expressions are recorded in Appendix~\ref{app:formulae}.

\section{Anchoring the low scale}
\label{sec:Higgs-stability}

We now turn to the scalar sector of the model, in particular the mechanism by which a set of light electroweak doublets 
acquire their non-zero vevs (from which we identify the SM Higgs). By requiring limited tuning in this Higgs sector, 
we will place constraints on the various symmetry breaking scales of the model, thereby anchoring the 
masses of the lightest new physics particles close to the TeV scale.

\subsection{The light-Higgs sector}

The theory contains several scalar fields that, at the end of the breaking chain in Fig.~\ref{fig:SSB}, transform as doublets of 
$SU(2)_L$ and hence as SM Higgs fields. We denote as the ``light-Higgs sector'' the subset comprised of the four 
doublets $\{\H^{22\pm}_a\}$ (see Sect.~\ref{sect:Yuk_mix}), {\em i.e.} those with second-family flavour indices, 
plus the four doublets from $H_a$.  
As anticipated, these two sets mix because of the non-vanishing vev of the $\Sigma_{L,R}$ fields. More precisely, the doublets in $\H^{22\pm}_a$ mix with the doublets in $H_1$ but not $H_{15}$ (at least not at the renormalisable level), and this mixing happens separately for the up-type and down-type components, {\em i.e.}~only fields with the same $U(1)_R$ charges can mix. 

In this light-Higgs sector we expect a single `ultra-light' component to appear, that we identify with the 
effective SM-like Higgs. Without loss of generality, in order to simplify the notation,  
we assume that this ultra-light component can be identified with the up-type component of $H_1$.
 In other words, we assume
\be
|\langle H_1 \rangle|  \gg |\langle H_{15} \rangle| \qquad  \text{and} \qquad |\langle H^-_1 \rangle|  \gg |\langle H^+_{1} \rangle| \,.
\label{eq:SMH_conf}
\ee
where $|\langle H_a \rangle|$ denotes the magnitude of the vev of the field $H_a$.
To this end, we recall that it is sufficient that either $H^-_1$ or $H^+_{1}$ acquire a non-vanishing vev in order to 
achieve the spontaneous symmetry breaking of electroweak symmetry and give non-vanishing masses to all the fermions. 
Indeed, via the $\mathcal{R}_a$ and 
$\mathcal{K}_a$ rotation matrices introduced above, the vev of a single (up or down) component of $\H^{22}_a$ generates non-vanishing masses
for both up and down quarks (and a similar mechanism holds for third-generation quarks).
 A non-vanishing vev for $H_{15}$ is needed to generate the appropriate splitting between $y_\tau$ and  $y_b$.
 However, the smallness of $y_{\tau,b}$ with respect to $y_t$ ensures that this effect can easily be obtained with 
$|\langle  H_{15} \rangle| \ll  |\langle H_{1}\rangle|$, hence with a tiny contribution of $H_{15}$ to the electroweak vev
 (and, correspondingly, to the SM-like Higgs field). 

As we shall see in the next section, the configuration in (\ref{eq:SMH_conf}) can
be obtained with a rather natural choice of parameters in the effective Higgs potential. 
The only critical point is to ensure  $|\langle H_1 \rangle| \approx v$. This condition is what implies 
 $\Lambda_\Sigma  \lesssim 10$~TeV,  anchoring the whole 
chain of  symmetry breaking scales in  Fig.~\ref{fig:SSB}. 

\subsubsection{Mixing and spectrum in the light-Higgs sector} \label{sec:SigmaMixHiggs}

The mixing between the $\{\H^{22\pm}_a\}$ fields and the components from $H_1$ is induced by the following interaction terms,
\be
\cL \supset \lambda^1_{H\Sigma} \H_1^{22} \Sigma_L H_1 \Sigma_R + \lambda^{15}_{H\Sigma} \H_{15}^{22} \Sigma_L H_1 \Sigma_R + {\rm h.c.} \, ,
\label{eq:lambda_HSigma}
\ee
where we use the generic notation `$\H^{22}_a$' to denote $SU(3)_{1+2}$ singlet scalar components originating from the UV Higgses $\H_1$ and $\H_{15}$. Recall that under the light-flavour deconstruction, $Sp(4)_L \to SU(2)_{L,1}\times SU(2)_{L,2}$,
the $\Sigma_L$ field decomposes as
\be
\Sigma_L \sim {\bf 4} \to {\bf (2,1)} \oplus {\bf (1,2)}\, .
\ee
After $SU(2)_{L,1} \times SU(2)_{L,2} \to SU(2)_{L,1+2}$, this just gives two doublets of $SU(2)_{L,1+2}$ (see (\ref{eq:SigmaL_decomp})). 
We assume that the vev of $\Sigma_L$ occurs in the ${\bf (1,2)}$ component that couples to $\H^{2j}_a$, and likewise that the vev of $\Sigma_R$ occurs in the ${\bf 1}_{\pm 1/2}$ components of $SU(2)_{R,1} \times U(1)_R^{\prime\prime}$ that couple only to $\H^{i2}_a$.
This way, once $\Sigma_{L,R}$ acquire their vevs, the mixing term in (\ref{eq:lambda_HSigma})  selects only the $\H^{22}_a$ components.

We now focus the attention on the up-type $SU(4)$-singlet Higgs fields, {\em i.e.}~the fields that after $SU(2)_{L,1+2} \times SU(2)_{L,3} \to SU(2)_{L}$
behave as $SU(2)_{L}$ doublets with (SM) hypercharge $-1/2$. Their mass matrix, ${\bf M_u^2}$, defined via
\be
\cL \supset ({\vec H}^{-})^\dagger {\bf M_u^2} {\vec H}^-
\ee
where ${\vec H}^-=( \H^{22-}_1, \H^{22-}_{15},  H_1^-)$, assumes the form
\be {\bf M^2_u} = 
\begin{pmatrix}
(M^u_{1})^2& 0 &\lambda^1_{H\Sigma} |\langle \Sigma_L \rangle| |\langle \Sigma_R \rangle| \\
0&(M_{15}^u)^2&\lambda^{15}_{H\Sigma} |\langle \Sigma_L \rangle||\langle \Sigma_R \rangle| \\
\lambda^1_{H\Sigma} |\langle \Sigma_L \rangle||\langle \Sigma_R \rangle| 
& \lambda^{15}_{H\Sigma} |\langle \Sigma_L \rangle||\langle \Sigma_R \rangle| & m_u^2
\end{pmatrix}\,.
\ee
A completely analogous structure holds for  ${\bf M^2_d}$, acting on the vector $( \H^{22+}_1, \H^{22+}_{15},  H_1^+)$,
with identical off-diagonal entries and potentially different diagonal elements.\footnote{Generically, the components $\H_1^{22-}$ and $\H_{15}^{22-}$ are expected to mix and populate the 1-2 and 2-1 entries of the upper blocks of ${\bf M^2_{u,d}}$. We work under the assumption that this mixing is absent as its only effect would be to re-define effective couplings $\lambda^a_{H\Sigma}$, leaving the following discussion essentially unchanged.}

The mass matrix ${\bf M^2_u}$ can de diagonalized by an appropriate orthogonal 
rotation of the three fields,
\be
O_u {\bf M^2_u} O_u^T = {\rm diag} ( M^2_{u_1},  M^2_{u_{2}},  -\mu_u^2 )~. 
\ee
Assuming $m_u^2$,  $\lambda^a_{H\Sigma} |\langle \Sigma_{L,R} \rangle|^2 \ll  (M^u_{a})^2$, we find
\be
\mu_u^2 \approx  \sum_{a=1,15} (\theta_u^a)^2  (M^u_{a})^2 -  m_u^2\,,
\label{eq:mu2}
\ee
where
\be
 \theta_u^a = \lambda^a_{H\Sigma} \frac{  |\langle \Sigma_L \rangle|| \langle \Sigma_R \rangle|   }{ (M_{a}^u)^2 }\,, \qquad 
  O_u \approx \begin{pmatrix}  1 & 0  & \theta_u^{1} \\   0  & 1  & \theta_u^{15}  \\ - \theta_u^{1} &  - \theta_u^{15} & 1 \end{pmatrix}\,. 
\ee

Electroweak symmetry breaking is achieved if either ${\bf M^2_u}$ and/or ${\bf M^2_d}$ develops 
small negative eigenvalues. This requires some tuning between 
the two terms in (\ref{eq:mu2}), or the analogous combination for ${\bf M^2_d}$.
It is natural to assume this happens only in one the two mass matrices: we work under the hypothesis this 
happens only in ${\bf M^2_u}$.
In this limit,  the effective mixing parameters defined in  (\ref{eq:epsh}) assume the form
\be \label{eq:eta_hypothesis}
\epsilon_h = \sqrt{ (\theta_u^1)^2 + (\theta_u^{15})^2 }\,,
\qquad  
\eta^-_a  \approx \frac{\theta_u^{a}}{ \sqrt{ (\theta_u^1)^2 + (\theta_u^{15})^2 }  } \frac{ v_u}{v}\,,  \qquad 
\eta^+_{a}=0\,.
\ee
Here $v_u \propto \mu_u$ denotes the `fraction' of the electroweak-breaking vev that is due to the $H_1$ field, 
while $v$ denotes the total vev considering also the contribution from  $\langle H_{15} \rangle$. The latter is 
controlled by independent parameters and can easily be tuned to be small, reaching the configuration 
(\ref{eq:SMH_conf}).

\subsubsection{Constraints on the scales from the electroweak vacuum}
The requirement that $\mu^2_u \sim v^2$, with minimal tuning, allows us to derive a series of constraints 
on the symmetry breaking scales that appear in our setup. Assuming that both terms in (\ref{eq:mu2}) are of similar size 
implies 
\be
v^2 \sim \lambda^2_{H\Sigma} \frac{  \Lambda_\Sigma^4}{  M^2_\H } \quad \to \quad
\frac{  \Lambda^2_\Sigma}{ M_\H}  \sim \frac{ v}{ \lambda_{H\Sigma}}\,,
\label{eq:FT1}
\ee
where $\lambda_{H\Sigma}$ denotes generically $\lambda_{H\Sigma}^{1,15}$,
and $M_\H$ generically denotes $M^u_{1,2}$.
At the same time, the request $\epsilon_h \sim 10^{-2}$ from the Yukawa couplings implies 
\be
\epsilon_h \sim \lambda_{H\Sigma} \frac{ \Lambda_\Sigma^2 }{  M_\H^2 }   \quad \to \quad  \frac{M_\H}{  \Lambda_\Sigma}    \sim \left( \frac{\lambda_{H\Sigma}}{\epsilon_h} \right)^{1/2}\,.
\label{eq:FT2}
\ee
The $\lambda_{H\Sigma}$ couplings cannot be arbitrarily small, since  
the operators in (\ref{eq:lambda_HSigma}) are necessarily generated by quantum corrections: a natural lower bound is 
$|\lambda_{H\Sigma}| \gtrsim 10^{-2}$, which implies that $\Lambda_\Sigma$ is at most as large as $M_\H$.

The last piece of information we need to take into account are 
the experimental bounds on the family non-universal electroweak gauge bosons ($W^\prime, Z^\prime$) generated
by the breaking $SU(2)_{L,1+2} \times SU(2)_{L,3}\to SU(2)_L$. In particular from the stringent $Z^\prime \to \ell\bar\ell$ bounds~\cite{CMS:2021ctt}
we deduce  $\Lambda_\Sigma \gtrsim 10 \text{~TeV}$. 
All these constraint can be satisfied for $|\lambda_{H\Sigma}| \sim 0.01$ and 
\be
M_\H \sim \Lambda_\Sigma\sim \text{~10~TeV}\,.
\ee
The range for $M_\H$ is well compatible with present bounds on heavy Higgs fields.

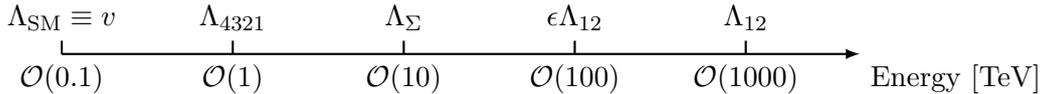
\begin{figure}[t]
\centering
\begin{tikzpicture}[x=1.5cm]
\draw[black,->,thick,>=latex]
  (0,0) -- (7,0) node[below right] {Energy [TeV]};
{
  \draw[black,thick] 
    (0,0) -- ++(0,5pt) node[above] {$\Lambda_\text{SM}\equiv v$};
  \draw[black,thick] 
    (1.5,0) -- ++(0,5pt) node[above] {$\Lambda_{4321}$};
  \draw[black,thick] 
    (3,0) -- ++(0,5pt) node[above] {$\Lambda_\Sigma$};
  \draw[black,thick] 
    (4.5,0) -- ++(0,5pt) node[above] {$\epsilon\Lambda_{12}$};
  \draw[black,thick] 
    (6,0) -- ++(0,5pt) node[above] {$\Lambda_{12}$};
}
\node[below,align=left,anchor=north,inner xsep=0pt] 
  at (0,0) 
  {$\mathcal{O}(0.1)$};  
\node[below,align=left,anchor=north,inner xsep=0pt] 
  at (1.5,0) 
  {$\mathcal{O}(1)$};  
\node[below,align=left,anchor=north,inner xsep=0pt] 
  at (3,0) 
  {$\mathcal{O}(10)$};   
\node[below,align=left,anchor=north,inner xsep=0pt] 
  at (4.5,0) 
  {$\mathcal{O}(100)$};  
\node[below,align=left,anchor=north,inner xsep=0pt] 
  at (6,0) 
  {$\mathcal{O}(1000)$};  
\end{tikzpicture}
\caption{\label{Fig:scales}The relevant energy scales in our model, which are associated with different symmetry breaking steps (as detailed in Fig.~\ref{fig:SSB}), are each separated by roughly an order of magnitude.}
\end{figure}

Combining these indications with the request $\epsilon_{L,R}\sim 10^{-1}$ from the light-fermion spectrum, and the bound $\Lambda_{12} \gtrsim 10^{3}$~TeV
 from flavour-changing processes involving the first two generations of quarks~\cite{Isidori:2010kg}, 
 we end up with a coherent spectrum where each of the four scales indicated 
 in  Fig.~\ref{fig:SSB} are separated by one order of magnitude: starting from $\Lambda_{12} \sim 10^{3}$~TeV down to $\Lambda_{4321} \sim 1$~TeV, as summarised in Fig.~\ref{Fig:scales}.
 On general grounds, the scalar sector of this framework is stable under quantum corrections if the scalars at a given scale receive 
 one-loop corrections only from scalars at the scale immediately above (and not from those at the higher scales)~\cite{Allwicher:2020esa}. 
 In our case, this condition is respected but for one exception, namely  $m_u^2$ ({\em i.e.}~the mass term of $H_1$), which could receive one-loop corrections from the $\Sigma_{L,R}$ fields. The natural expectation is thus $m_u \gtrsim 1$~TeV and not $m_u \lesssim v$. The model therefore requires some amount of 
 fine-tuning in order to reproduce the observed value of the electroweak scale. However, this tuning is not worse than that present in any 
 realistic SM extensions, given current bounds on direct searches for new physics. 
 This is the manifestation of the so-called {\em little--hierarchy} problem~\cite{Giudice:2017pzm} in our model.

\subsection{Leptoquark phenomenology} \label{flavano}
We are now ready to analyse some of the phenomenological implications of our model at both low and high energies.
As summarised in Table~\ref{tab:Matter_Content_4321},
the model posseses a rich spectrum of new states at the TeV scale. 
A detailed analysis of the possible signatures of all these states in high-energy $pp$ collisions 
is beyond the scope of this paper: these signatures vary a lot depending on  the mass matrices 
of the new sates which are poorly constrained by low-energy data.
On the contrary, rather precise predictions can be obtained for processes mediated by 
the TeV-scale $U_1$ leptoquark (LQ) generated by the $4321\to\,$SM breaking. 
Analysing these predictions is also interesting in order to compare the expectations of our model
with those of other 4321 completions.

In order to take advantage of a series of recent phenomenology analyses of 4321 models
(see e.g.~\cite{Cornella:2019hct,Cornella:2021sby,Aebischer:2022oqe}), it is convenient to write the effective  $U_1$ 
interactions with the SM fermions as
\be
\mathcal{L}^{\rm int}_{\rm eff} =  \frac{g_U}{ \sqrt{2} } J_U^\mu U_\mu~+~{\rm h.c.}\, \qquad 
  J_U^\mu = \left( \beta_L^{i\alpha} \overline{Q}^i_L\gamma^\mu L^\alpha_L+\beta_R\,\overline{b}_R\gamma^\mu \tau_R \right)\,,
\label{eq:Jmu-down}
\ee
introducing the effective couplings $g_U$, $\beta_L^{i\alpha}$ and $\beta_R$. 
By convention, the quark doublet $Q^i$ is written 
in the down-quark mass-eigenstate basis ($i=b,s,d$), the lepton doublet $L^\alpha_L$ 
is written in the charged-lepton mass-eigenstate basis ($\alpha=e,\mu,\tau$), and 
$\beta_L^{b\tau}\equiv 1$.
Within our model, in the limit $\epsilon_h \to 0$, {\em i.e.}~neglecting the mixing of the light 
families with the third, we have $g_U=g_4$,  $|\beta_R| = 1$, and all the $\beta_L^{i\alpha}$ vanish but for 
$\beta_L^{b\tau}$.

As in all 4321 models, a key constraint on the flavour structure of the theory comes from the effective four-quark operators mediated 
by the TeV-scale color octet (coloron), which is also generated by the $4321\to\,$SM breaking. These effective   operators
 are strongly constrained by $B_s$--$\bar B_s$ and $B_d$--$\bar B_d$ mixing (see e.g.~\cite{Cornella:2021sby,Aebischer:2022oqe}). Satisfying these bounds 
requires the alignment of the third generation in the down sector, {\em i.e.}~identifying the left-handed quark doublet charged 
under $SU(4)$ as 
\be
 Q_L^3 \approx \left( \begin{array}{c} \sum_{q=u,c,t} V^*_{ub} q_L \\  b_L \end{array} \right)\,. 
\ee
This justifies, {\em a posteriori}, the choice of the  down-mass eigenstate basis 
in Eq.~(\ref{eq:Jmu-down}). 
Recall that, in our framework, this alignment condition follows from the hierarchy $\delta_d \ll \delta_u$, as in Eq.~(\ref{eq:deltas}) above, which is a natural consequence of assuming that the 4321-breaking vev occurred in the component labelled $\omega^{cb}$, as in Eq.~(\ref{eq:omega_choice}), given also that $|\langle \H^{22+}_a \rangle|=0$.  This implies that the heavy$\to$light mixing in the CKM matrix originates from the up-quark sector and that $\beta_L^{s\tau}, \beta_L^{d\tau}$ remain vanishing small, at the tree level, even 
if $\epsilon_h \not= 0$. 

We emphasize that, in contrast to other 4321 models in the literature, the required down-alignment of the quark Yukawa matrices does not have to be imposed by hand in this model, but follows from a natural symmetry breaking structure that matches the deeper UV dynamics onto 4321.

\subsubsection{$R_{D^{(*)}}$ and $pp \to \tau^+\tau^- + X$}
Integrating out the LQ at tree-level leads to the following effective Lagrangian relevant to $b\to c \ell\nu$ decays:
\begin{align}
   \cL_{b\to c}
  = - \frac{4 G_F}{\sqrt{2}} V_{cb} & \bigg[
    \Big( 1 + \cC_{LL}^{c} \Big)
    (\bar c_L \gamma_\mu b_L) (\bar\tau_L \gamma^\mu \nu_L)  - 2\,\cC_{LR}^{c} \,
    (\bar c_L b_R) (\bar\tau_R\,\nu_L) \bigg] \,,
    \label{eq:Cc}
\end{align} 
where 
 \bea
  \cC_{LL}^c = \frac{ v^2 }{2  \Lambda_U^2 }  \left( 1 +\beta_L^{s\tau}\frac{V_{cs}}{V_{cb}}\right)\,,  \qquad  \cC_{LR}^c =  \beta^*_R  \, \cC_{LL}^c\,,
 \label{eq:charmWCs}
\eea
where we have written these Wilson coefficients in terms of the effective couplings introduced  in Eq.~(\ref{eq:Jmu-down}). The parameters $\cC_{LL(R)}^c$ can be extracted from data via 
the lepton flavour  universality (LFU) ratios $R_{D^{(*)}}$ using the following phenomenological expressions~\cite{Cornella:2021sby}: 
\begin{align}
 \Delta R_D  \equiv    \frac{R_D}{R_D^{\text{SM}}} -1 =  \,\, &  {\rm Re} \left( 2\, \cC^c_{LL} -3.00\,  \cC^{c\,*}_{LR} \right) +\cO[(\cC_{LL(R)}^c)^2] \,,  \no \\
 \Delta R_{D^*} \equiv    \frac{R_{D^*}}{R_{D^*}^{\text{SM}}} - 1 =  \,\,  & {\rm Re} \left( 2\, \cC^c_{LL} -0.24\, \cC^{c\,*}_{LR}  \right)+\cO[(\cC_{LL(R)}^c)^2] \,. 
    \label{eq:RDRds}
\end{align}
According to the recent analysis of $b\to c \ell\nu$  data in~\cite{Aebischer:2022oqe}, if $|\beta_R| = 1$ (as expected in our model) 
a very good fit to present data  is obtained for $\beta_R=-1$, a phase choice that maximises the interference of left- and right-handed currents 
in Eq.~(\ref{eq:RDRds}).\footnote{The phase of $\beta_R$ is a free parameter in our model determined by the relative sign of 
$M_{33}^d$ and $M_{33}^e$.}
The correspondingly preferred value of $\cC_{LL}^c$ is 
\be
\left. \cC_{LL}^c \right|^{\rm exp}_{\beta_R =-1} = 0.03 \pm 0.01\,.
\ee

The maximal value of $\cC_{LL}^c$ in the model is determined by the experimental lower bound on $M_U/g_U$ 
extracted from  high-energy data. For large LQ masses, the high-energy process $pp \to \tau^+\tau^- + X$,
to which the LQ contributes via the $t$-channel exchange, turns out to be the most effective probe. 
From the recent CMS analysis in~\cite{CMS:2022zks}, focused on the LQ $t$-channel exchange amplitude, 
one extracts the bound $\Lambda_U  \gtrsim 1.6~{\rm TeV}$~\cite{Aebischer:2022oqe,Haisch:2022afh}, with a tantalizing $3\sigma$ excess for $\Lambda_U  \approx 1.6~{\rm TeV}$.
Setting  $\Lambda_U = 1.6~{\rm TeV}$ and $\beta_L^{s\tau}=0$ we get 
$\left. \cC_{LL}^c \right|^{\rm th}_{\beta_R =-1} = 0.012$ or, equivalently,  
\be
\Delta R_D \approx  6.0\%\,, \qquad   \Delta R_{D^*} \approx  2.7\%\,.
\ee 
These values are already within the $2\sigma$ range determined by 
low-energy data, implying a significant agreement compared to the SM expectations in these obseravables.
A further improvement could be obtained with a value of $\beta_L^{s\tau}\approx0.01$, which could be generated beyond the tree level (see the next Section).
In this case $\left. \cC_{LL}^c \right|^{\rm th}_{\beta_R =-1}$  could raise up to $\approx 0.015$, which is well within the 90\%CL 
experimental range.

\subsubsection{$R_{K^{(\ast)}}$}
The LQ effective  interaction in Eq.~(\ref{eq:Jmu-down}) also leads to a tree-level contribution to $b\to s\mu^+\mu^-$ amplitudes
that, in turn, induce non-vanishing corrections to the neutral-current LFU ratios $R_{K}$ and $R_{K^*}$. 
However, as we shall see, these effects are naturally quite small in our setup.

Adopting the standard convention to define the $b\to s\mu^+\mu^-$ effective operators $O_{9,10}$ (see e.g.~\cite{London:2021lfn}),
we get
\be
\Delta C_{9}^{\mu \mu} =- \Delta C_{10}^{\mu \mu} = -   \frac{ v^2 }{ 2 \Lambda_U^2 } \left(
\frac{  2 \pi }{ \alpha_{\rm em} }  \frac{  \beta_L^{s \mu} \beta_L^{b \mu*}  }{   V_{t b} V_{t s}^{*} } \right)\,.
\ee
In terms of these modified Wilson coefficients, the LFU ratio $R_K$ in the dilepton mass
interval  $m^2_{\ell\ell} \in[ 1~{\rm GeV}^2, 6~{\rm GeV}^2]$
reads~\cite{Cornella:2021sby}: 
\be
\Delta R_K =  R_K - 1 =  0.50\, \left. {\rm Re} \left(\Delta C_{9}^{\mu \mu}\right)  \right|_{ \Lambda_U = 1.6~{\rm TeV} }  
\approx  0.13 \times 
{\rm Re} \left( \frac{  \beta_L^{s \mu} \beta_L^{b \mu*}  }{   10^{-3} } \right)\,,
\ee
and in the same $m^2_{\ell\ell}$ interval we have $\Delta R_{K^*} \approx \Delta R_K$.

\begin{figure} [t]
  \centering
\includegraphics[width=0.9\textwidth]{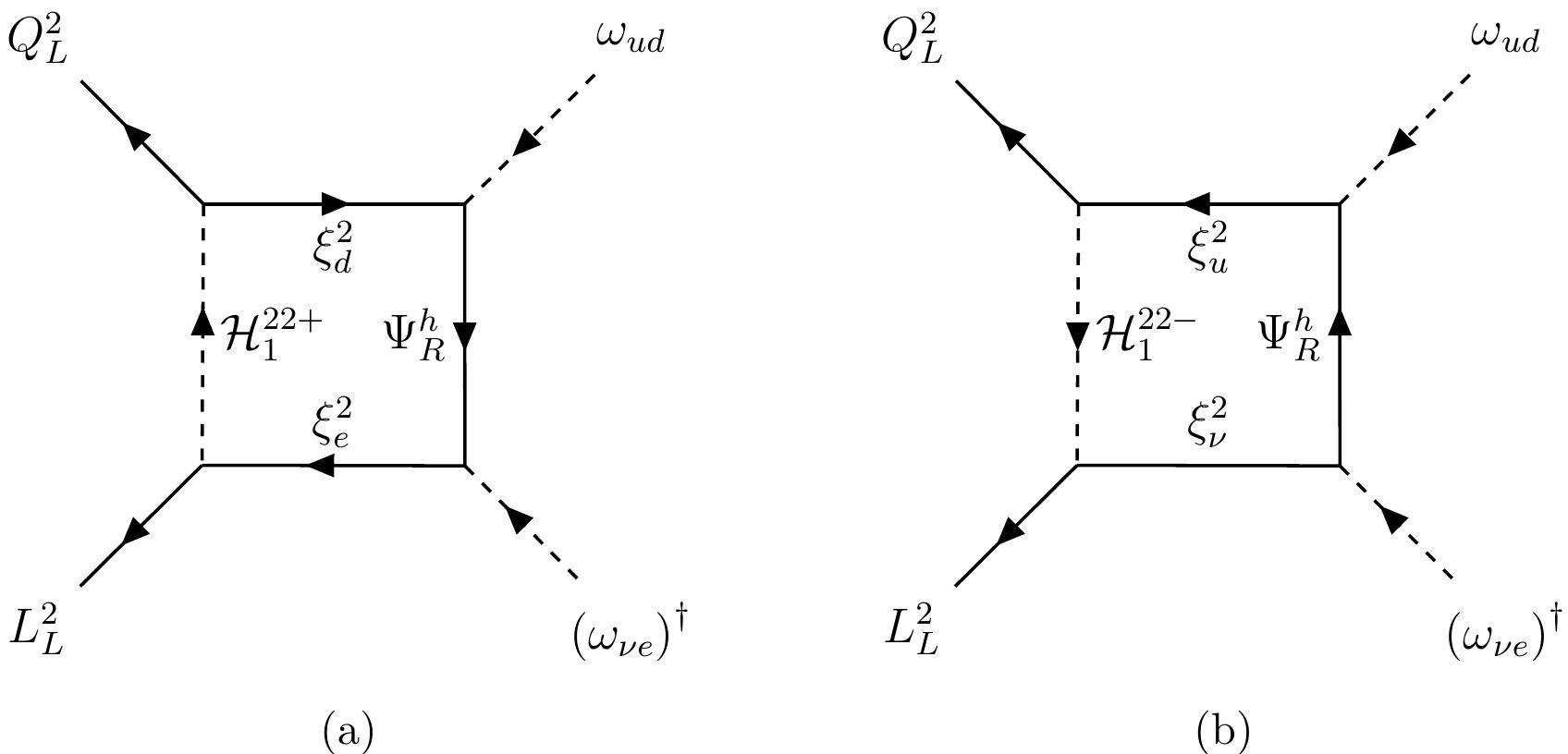}
\caption{Box diagrams leading to an effective $\bar Q_L^2 \gamma^\mu L_L^2 U_\mu$ interaction,
induced by the exchange of a (heavy) $SU(3)^l$--singlet Higgs. \label{fig:rk1}}
  \end{figure}
\begin{figure} [t]
  \centering
\includegraphics[width=0.9\textwidth]{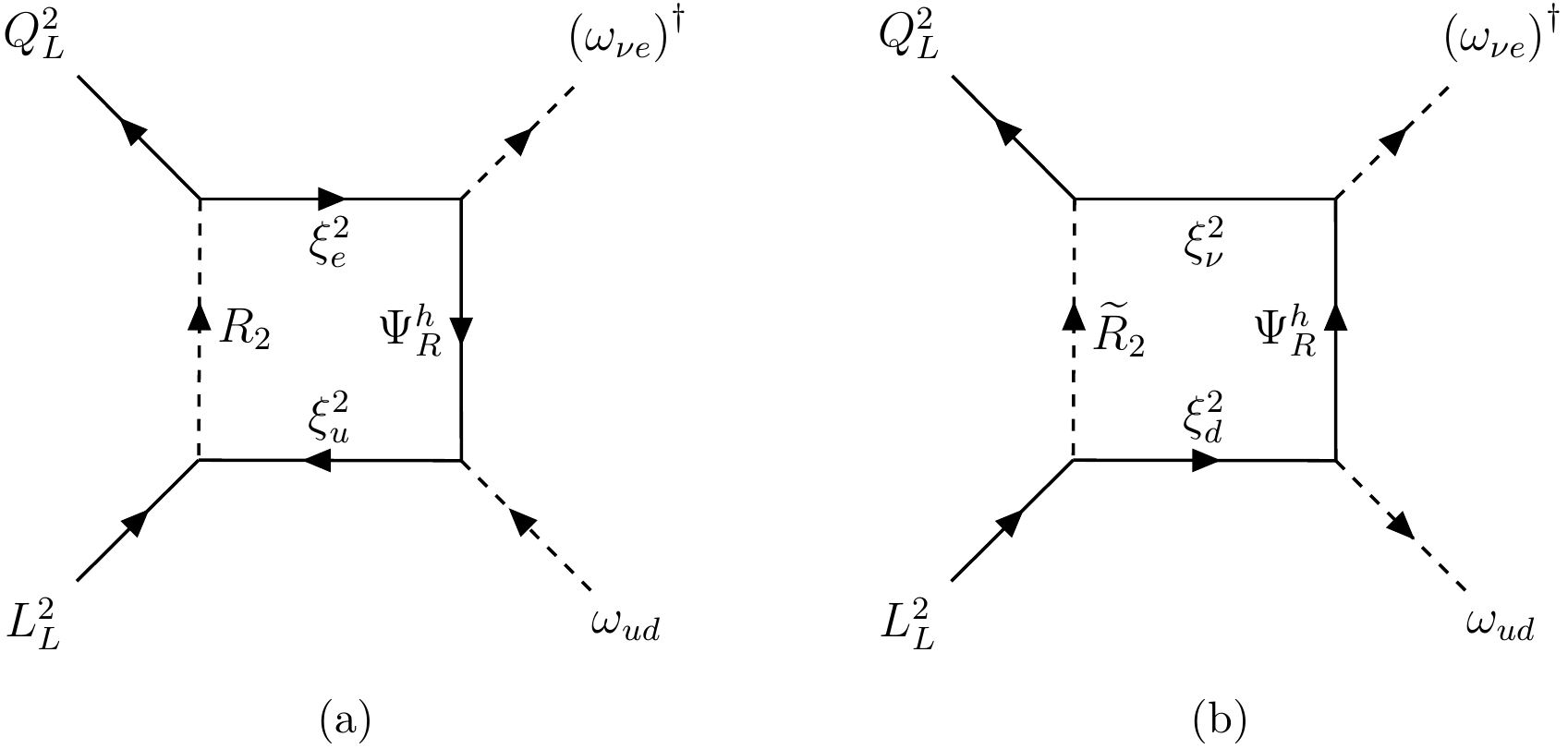}
\caption{Box diagrams leading to an effective $\bar Q_L^2 \gamma^\mu L_L^2 U_\mu$ interaction, induced by the 
exchange of a (heavy)  $SU(3)^l$--triplet Higgs  (these components originate from the decomposition of $\mathcal{H}_{15}$). \label{fig:rk2}}
  \end{figure}

Within our model, a non-vanishing $\beta_L^{b\mu}$ is generated by the diagonalization of the lepton mass matrix, 
\begin{align}
| \beta_L^{b\mu} | \approx\left | \frac{\cM^e_{23} }{ \M_{33}^e }  \right| = O(\epsilon_h \delta_e )  < 10^{-1}\,.
\end{align}
On the contrary,  the diagonalization of the quark mass matrices do not lead to a  non-vanishing $\beta_L^{s\mu}$ 
in the limit of perfect down alignment. 

An effective coupling of the LQ to second generation fermions 
is generated in the model beyond the tree level by the one-loop diagrams shown in Figs.~\ref{fig:rk1} and \ref{fig:rk2}.
These one-loop diagrams generate a dimension-six operator of the type 
$\bar Q_L^i \gamma^\mu L_L^\alpha \omega^{i 3}_{ud} D_\mu (\omega_{\nu e}^{3 \alpha})^\dagger$ 
that, once the $\omega$ fields acquire their vevs, leads to a non-vanishing 
$\bar Q_L^2 \gamma^\mu L_L^2 U_\mu$ effective interaction. 
The  $\beta_L^{s\mu}$  thus generated can be written as 
\begin{align}
\beta_L^{s\mu} =  \zeta \frac{\M_{22}^e \M_{22}^{d\ast}}{|\M_{22}^e ||\M_{22}^{d}|}\,,
\label{eq:betasm}
\end{align}
where 
\be
\zeta \equiv \frac{\kappa_{1,15}^2 \lambda_q \lambda_\ell v_{3}v_{1}}{16\sqrt{2}\pi^2}\sum_{\text{box diag.}} \Phi[M_{\H},m_{\xi_\ell},m_{\xi_q}]\,,
\ee
with $\lambda_{q,\ell}$ denoting the couplings between $\omega$, $\Psi^h_R$, and $\Xi$, 
and where the loop function is
\be
\Phi[M_{\H},m_{\xi_\ell},m_{\xi_q}] \equiv - \frac{M_{\H}^2-m_{\xi_\ell}m_{\xi_q} +m_{\xi_\ell}m_{\xi_q} \log \left(\frac{m_{\xi_\ell}m_{\xi_q}}{M_{\H}^2} \right)}{\left(m_{\xi_\ell}m_{\xi_q}-M_{\H}^2 \right)^2}\,.
\label{formphi}
\ee
The sum over all relevant box diagrams yields
\begin{align}
\zeta = \frac{\lambda_q \lambda_\ell v_{3}v_{1}}{16\sqrt{2}\pi^2}\Big(&\kappa_1^2\Phi[M_{\mathcal{H}^{22+}_1},m_{\xi_e},m_{\xi_d}]+\kappa_{15}^2\Phi[M_{\mathcal{H}^{22+}_{15}},m_{\xi_e},m_{\xi_d}] \\
&+\kappa_1^2\Phi[M_{\mathcal{H}^{22-}_1},m_{\xi_\nu},m_{\xi_u}]+\kappa_{15}^2\Phi[M_{\mathcal{H}^{22-}_{15}},m_{\xi_\nu},m_{\xi_u}] \nonumber\\
&+\kappa_{15}^2\Phi[M_{R_2},m_{\xi_e},m_{\xi_u}]+\kappa_{15}^2\Phi[M_{\widetilde{R}_2},m_{\xi_\nu},m_{\xi_d}]\nonumber \\
&+ \kappa_a \rightarrow \bar{\kappa}_a \Big)\,. \nonumber
\end{align}
The expression for $\beta_L^{s\mu}$ in (\ref{eq:betasm}) is suppressed both by the loop factor $1/(16\pi^2)$ and by the 
scale ratio $v_{1,3}^2 / M^2_\H$. If all the heavy scalars running inside the loops
have masses of $\mathcal{O}(10~{\rm TeV})$, as expected given that the corresponding  
scalar potential is characterised by the scale $\Lambda_\Sigma$, 
the corrections to $R_{K^{(*)}}$ should not exceed $1\%$. On general grounds, we thus expect tiny modifications to $R_{K^{(*)}}$ 
below the detectable level, consistent with the recent findings of the LHCb Collaboration~\cite{LHCb:2022zom}.

In specific regions of parameter space, mild cancellations in the effective potential could bring some of the 
scalar masses down to the TeV range, leading to larger corrections to $R_{K^{(\ast)}}$, close to the maximal values allowed by present data. 
For instance, a consistent benchmark point leading to 
$\Delta C_{9}^{\mu \mu} \sim -0.1$ (hence $\Delta R_K \approx - 5\%$) is obtained for
\begin{align}
&m_{\xi_q^u} \sim 1.4 \text { TeV }, \quad m_{\xi_q^e} \sim 1.1 \text { TeV } , \quad v_{3} \sim 1.7 \text { TeV } , \quad v_{1} \sim 1.7 \text { TeV },
\no \\
&m_{\mathcal{H}_1^{22\pm}} \sim 12.3 \text { TeV }, \quad m_{\mathcal{H}_{15}^{22\pm}} \sim 12.5 \text { TeV }, \quad m_{R_2,\widetilde{R}_2} \sim 1.1 \text { TeV }\,.
\end{align}
We emphasize that, to achieve such a shift in $C_{9}^{\mu \mu}$, the masses of the $R_2$ and $\widetilde{R}_2$ scalar leptoquarks are an order of magnitude lighter than their `natural' size of $\mathcal{O}(10~{\rm TeV})$.

\section{Conclusion}\label{conclu}

In this paper we explore a new kind of gauge model for explaining the origin of flavour in the Standard Model (SM). We suppose that physics in the UV is described by gauge interactions that unify quarks and leptons but in a fundamentally flavour non-universal way. We invoke a ``2+1'' family structure {\em i.e.} with the third family coupling to its own set of UV gauge bosons. On the other hand, within the first two generations we employ the rigid structure of electroweak-flavour unification. This unifying symmetry naturally controls flavour-changing processes in the 1-2 sector to be small, while generating hierarchical masses and mixing angles in the 1-2 sector through symmetry breaking steps at high energy scales. At lower energies of $\mathcal{O}(10 \text{~TeV})$, the light- and heavy-sectors are linked together, to match onto a non-universal `4321 model', which is finally broken to the SM near the TeV scale.

A model of this kind, while seemingly quite complex, is well motivated on general grounds by our current knowledge of particle physics in both the electroweak and flavour sectors. In summary, it naturally explains:
\begin{itemize}\setlength\itemsep{0.3em}
\item the spectrum of quarks and leptons and their seemingly {\em ad hoc} pattern of hypercharges, within each SM family, via enlarged $SU(4)$ colour symmetries;
\item the observed hierarchical pattern of fermion masses and quark mixing angles, with $\mathcal{O}(1)$ Yukawa couplings for the third family;
\item why there exist two generations that are `light' {\em i.e.} with suppressed Yukawa couplings $\ll 1$, that are in this sense similar from the point of view of the Higgs sector;
\item why flavour-changing transitions in this 1-2 sector are consistent with the SM, probing new physics contributions up to very high (effective) scales;
\item why a SM Higgs boson of $\mathcal{O}(0.1\text{~TeV})$ mass is not unnatural (beyond an unavoidable tuning of about 1 part in 100 as per the `little hierarchy problem'), given our ladder of symmetry breaking scales can be anchored at the TeV scale.
\end{itemize}
The output of such a model-building framework, from the phenomenological perspective, is a new physics sector coupled dominantly to the third generation fermions, in the vicinity of the TeV scale. 

We have not comprehensively studied the phenomenology of the new physics particles in this work, but rather pointed out the most obvious effects.
One of the lightest new particles is a $U_1$ vector leptoquark with flavour non-universal couplings. As in other 4321 models, the leptoquark parameter space is cornered by high-$p_T$ data and complementary constraints on the coloron and $Z^\prime$ gauge bosons that necessarily accompany it. 
An interesting aspect of this model, compared to generic 4321 constructions, is the 
possibility to justify the down-alignment of the heavy $SU(4)$
gauge bosons --phenomenologically required to satisfy the tight $\Delta F=2$ bounds-- as a result of the vacuum structure of the 
link fields mediating $4321 \to  {\rm SM}$ breaking. These link fields carry remnant light-flavour indices as a result of the high-scale 
electroweak-flavour unification. 

We discussed the extent to which the $U_1$ leptoquark
can explain the hints of new physics in charged-current ($b \to c \tau \nu$) $B$-meson decays. The model can naturally accommodate a 6\% and 3\% increase in $R_{D}$ and  $R_{D^{(\ast)}}$ respectively. While not matching the current central values of these observables, these effects significantly ease the tension with respect to the SM predictions.
The relative impact on $b \to s \mu\mu$ amplitudes is naturally smaller, except for very specific regions 
of the parameter range where it can reach at most the few \% level, consistent with recent findings~\cite{LHCb:2022zom}.

Beyond these particles, there are other states near the TeV scale that differ from other 4321 models, most notably in the scalar sector. For example, we have a suite of heavy scalar particles in the same representation as the SM Higgs, but with heavier masses and with $\mathcal{O}(1)$ Yukawa couplings to the second generation fermions. There is also a set of vector-like fermions, needed to generate the CKM rotation angles involving the third family, which are also expected to be rather light (certainly if the $U_1$ leptoquark is to mediate any appreciable contribution to $b \to s \mu\mu$ transitions). We save phenomenological explorations of the high-$p_T$ signatures of these particles for future work.

\acknowledgments 

JD is grateful to J. Tooby-Smith for discussions and for collaboration on related work.
This work is funded by the European Research Council (ERC) under the European Union’s Horizon 2020 research and innovation programme under grant agreement 833280 (FLAY), and by the Swiss National Science Foundation (SNF) under contract 200020-204428.

\appendix
\section{Formulae for fermion masses and mixing angles} \label{app:formulae}

In this Appendix we record the precise formulae for the quark and charged lepton masses and mixings in our model, in terms of the fundamental UV couplings of the theory, obtained by tree-level EFT matching at each symmetry breaking step.

We begin by writing the mass matrices $M^f$ for each type of fermion, $f \in \{u,d,e\}$. There are contributions from both the `light-Higgs sector' fields $\{\H_a^{22\pm}\}$, which recall get vevs $|\langle \H_a^{22\pm} \rangle| = \epsilon_h \eta_a^\pm v$, and from $\{H_a\}$, for which we parametrize the vevs as
\begin{align}
\langle H_1 \rangle &= v_1^{-} B_1 \otimes C_2 - v_1^{+} B_2 \otimes C_1\, , \\
\langle H_{15} \rangle &= \left(\sum_{i=1}^3 A_i \otimes A_i^\ast - 3A_4 \otimes A_4^\ast \right) \otimes \left(v_{15}^{-} B_1 \otimes C_2 - v_{15}^{+} B_2 \otimes C_1 \right)\, ,
\end{align}
using the bases defined in \S \ref{sec:conventions} of the main text. While in the main text we adopt the hypothesis that $\eta_a^+=0$, as given by Eq.~(\ref{eq:eta_hypothesis}), in this Appendix we give general formulae valid for any $\eta_a^{\pm}$.

Each mass matrix can be decomposed into the following `2+1' block structure
\begin{align}
\sqrt{2} M^f = 
\begin{pmatrix}
\epsilon_h \widehat{\mathcal{M}}^f & \epsilon_h \delta_f \widehat{\bf n}^f \\
{\bf 0}^T & M_{33}^f
\end{pmatrix}\, ,
\end{align}
where recall the $\delta_f$ parameters are defined in Eq. (\ref{eq:deltas}).
Defining $\Gamma_1 = 1$ and $\Gamma_{15} = -3$, which encodes the relative values of the Higgs vevs on the charged leptons, and summing over $a \in \{1,15\}$, we have
\begin{align}
M_{33}^u &= y_a^3v_a^- +\overline{y}_a^3 v_a^{+\ast}  \, ,\\
M_{33}^d &= y_a^3 v_a^+ +\overline{y}_a^3 v_a^{-\ast}\, ,\\
M_{33}^e &=\Gamma_a [y_a^3 v_a^+ + \overline{y}_a^3 v_a^{-\ast}\ ] \, .
\end{align}

The $U(2)$-breaking vectors $\widehat{\bf n}^f$ that mix the light families with the third are each given by
\be
\widehat{\bf n}^f = 
\begin{pmatrix}
\epsilon_L & 0 \\
0 & 1
\end{pmatrix}
{\bf n}^f\, ,
\ee
where
\begin{align}
{\bf n}^u &= v \big(\, \beta_L^a (\kappa_a \eta_a^- + \overline{\kappa}_a \eta_a^{+\ast}),\quad (\kappa_a \eta_a^- + \overline{\kappa}_a \eta_a^{+\ast})\,\big)^T = (n_1^u, \, n_2^u)^T\, , \\
{\bf n}^d &= v \big(\, \beta_L^a (\kappa_a \eta_a^+ + \overline{\kappa}_a \eta_a^{-\ast}),\quad (\kappa_a \eta_a^+ + \overline{\kappa}_a \eta_a^{-\ast})\,\big)^T=  (n_1^d, \, n_2^d)^T\, , \\
{\bf n}^e &= v \big(\, \beta_L^a \Gamma_a (\kappa_a \eta_a^+ + \overline{\kappa}_a \eta_a^{-\ast}),\quad  \Gamma_a(\kappa_a \eta_a^+ + \overline{\kappa}_a \eta_a^{-\ast})\,\big)^T= (n_1^e, \, n_2^e)^T\, ,
\end{align}
still summing over $a \in \{1,15\}$ in each term (the relative weighting by $\beta^{1,15}_L$ means that the first component of these vectors is not simply a rescaling of the second component by the same factor).

Finally, we define the upper-left 2-by-2 blocks, whose structures are generated by the EWFU mechanism. It is convenient to pull out the overall $\epsilon_{L,R}$ dependence, by defining 
\begin{equation}
\widehat{\mathcal{M}}^f = 
\begin{pmatrix}
\epsilon_L \epsilon_R {\mathcal{M}}^f_{11} & \epsilon_L {\mathcal{M}}^f_{12} \\
\epsilon_R {\mathcal{M}}^f_{21} &  {\mathcal{M}}^f_{22}
\end{pmatrix}\, .
\end{equation}
Then the `reduced' matrix elements are
\begin{align}
{\mathcal{M}}^u_{11} &= v \left[(\beta_{LR}^a + 2 \beta_L^a \beta_R^a)z_{+} + (\beta_{LR}^\ast +2 \beta_L^a \beta_R^{a \ast}) z_{-}^\ast )\right] \left( y_a^l \eta_a^- + \overline{y}_a^l \eta_a^{+\ast}\right)\, , \\
{\mathcal{M}}^u_{12} &= v \,\beta_L^a \left( y_a^l \eta_a^- + \overline{y}_a^l \eta_a^{+\ast}\right)\, , \\
{\mathcal{M}}^u_{21} &= v \left(z_{+} \beta_R^a + z_{-} \beta_R^{a \ast} \right) \left( y_a^l \eta_a^- + \overline{y}_a^l \eta_a^{+\ast}\right)\, , \\
{\mathcal{M}}^u_{22} &= v  \left( y_a^l \eta_a^- + \overline{y}_a^l \eta_a^{+\ast}\right) \, .
\end{align}
To obtain the corresponding formulae for $\mathcal{M}^d$, simply replace $y_a^l \eta_a^- + \overline{y}_a^l \eta_a^{+\ast} $ by $y_a^l \eta_a^+ + \overline{y}_a^l \eta_a^{-\ast}$ everywhere. To obtain the formulae for $\mathcal{M}^e$, additionally insert factors of $\Gamma_a$.

Using matrix perturbation theory, we calculate the eigenvalues of these matrices, to give the mass formulae:
\begin{align}
m_t &= |M_{33}^u|  \,, \\
m_b &= |M_{33}^d|  \,, \\
m_\tau &=|M_{33}^e|  \,, \\
m_c &= \epsilon_h |\mathcal{M}_{22}^u| \,, \\
m_s &= \epsilon_h |\mathcal{M}_{22}^d| \,, \\
m_\mu &= \epsilon_h |\mathcal{M}_{22}^e| \,, \\
m_u &= \epsilon_h \epsilon_L \epsilon_R \frac{|\mathrm{det}(\mathcal{M}^u)|}{|\mathcal{M}_{22}^u|}\, , \\
m_d &= \epsilon_h \epsilon_L \epsilon_R \frac{|\mathrm{det}(\mathcal{M}^d)|}{|\mathcal{M}_{22}^d|}\, , \\
m_e &= \epsilon_h \epsilon_L \epsilon_R \frac{|\mathrm{det}(\mathcal{M}^e)|}{|\mathcal{M}_{22}^e|}\, .
\end{align}
As described in the main text, the rough hierarchies are, in terms of our small model parameters $\epsilon_{L,R,h}$, given by $m_2/m_3 \sim \epsilon_h$ and $m_1/m_2 \sim \epsilon_L \epsilon_R$.

The CKM matrix is a little more involved. We have, firstly, the unsuppressed CKM elements on the leading diagonal:
\begin{align}
V_{ud} &= \frac{\M_{22}^{d\ast}\M_{22}^u \det(\M^d \M^{u\ast})}{\left|\M^d_{22} \M^u_{22} \det(\M^d \M^u)\right|}\, , \\
V_{cs} &= \frac{\M^d_{22}\M^{u\ast}_{22}}{\left|\M^d_{22} \M^u_{22}\right|}\, , \\
V_{tb} &=\frac{M^u_{33}M^{d\ast}_{33}}{\left|M^u_{33} M^d_{33}\right|}\, .
\end{align}
The next largest elements are the Cabibbo-suppressed CKM elements mixing the first and second generations,
where we emphasise the suppression by $\epsilon_L \sim \lambda$ in \textcolor{blue}{blue},
\begin{align}
V_{us} &= \frac{1}{\left|\M^d_{22} \M^u_{22}\right|}\left(\M^d_{12}\M^u_{22}\frac{\det(\M^{u\ast})}{\left|\det(\M^u)\right|}-\M^d_{22}\M^u_{12}\right)\textcolor{blue}{\epsilon_L}\, , \\
V_{cd} &=\frac{1}{\left|\M^d_{22} \M^u_{22}\right|}\left(\M^{u\ast}_{12}\M^{d\ast}_{22}\frac{\det(\M^{d})}{\left|\det(\M^d)\right|}-\M^{u\ast}_{22}\M^{d\ast}_{12}\right)\textcolor{blue}{\epsilon_L}\, .
\end{align}
Next, the CKM elements mixing the second and third generation are
\begin{align}
V_{cb} &= \frac{1}{\left|M^d_{33}\M^u_{22}M^u_{33}\right|} \left(n^d_{2}\M^{u\ast}_{22}\left|M^u_{33}\right| \textcolor{red}{\delta_d} - M^d_{33}  n^{u}_{2}\left|\M^u_{22}\right|\textcolor{red}{\delta_u} \right)\, , \\
V_{ts} &= \frac{1}{\left|M^d_{33}\M^d_{22}M^u_{33}\right|} \left(n^{u\ast}_{2}\M^{d}_{22}\left|M^d_{33}\right| \textcolor{red}{\delta_u} - M^{u\ast}_{33}n^{d\ast}_{2}\left|\M^d_{22}\right|\textcolor{red}{\delta_d} \right)\, ,
\end{align}
where the suppression by factors of $\delta_{u,d} \sim \lambda^2$ are denoted in \textcolor{red}{red}. Finally, we have the CKM elements mixing the first and third family,
\begin{align}
V_{ub}&=\frac{1}{\left|M^d_{33}\right|}\left(\frac{n^d_{1}\M^u_{22}}{\left|\M_{22}^u\right|}\frac{\det(\M^{u\ast})}{\left|\det(M^u)\right|} \textcolor{red}{\delta_d} -\frac{n^d_{2}\M^u_{12}}{\left|\M_{22}^u\right|} \textcolor{red}{\delta_d} - \frac{M^d_{33}n^u_{1}}{\left|M^u_{33}\right|} \textcolor{red}{\delta_u} \right)\textcolor{blue}{\epsilon_L}\, , \\
V_{td} &= \frac{1}{\left|M^u_{33}\right|}\left(\frac{n^{u\ast}_{1}\M^{d^\ast}_{22}}{\left|\M_{22}^d\right|}\frac{\det(\M^{d})}{\left|\det(M^d)\right|} \textcolor{red}{\delta_u} -\frac{n^{u\ast}_{2}\M^{d\ast}_{12}}{\left|\M_{22}^d\right|} \textcolor{red}{\delta_u} - \frac{M^{u\ast}_{33}n^{d\ast}_{1}}{\left|M^d_{33}\right|} \textcolor{red}{\delta_d} \right)\textcolor{blue}{\epsilon_L}\, ,
\end{align}
which, as expected, are doubly-suppressed.

For completeness, we conclude this Appendix by giving a formula for the Jarlskog invariant $J=\text{Im}\left[V_{us} V_{cb} V_{ub}^\ast V_{cs}^\ast \right]$, which captures the $CP$-violating phase in the CKM matrix.
First defining
\begin{align}
\mathcal{J}_1 &\equiv \left(\M^d_{22} \M^u_{12} |\det \left(\M^u\right)|-\M_{12}^d \M^u_{22} \det \left(\M^{u\ast}\right) \right) \, ,\\
\mathcal{J}_2 &\equiv \Big[|M^u_{33}|n^{d\ast}_{1} \M^{u\ast}_{22} \det \left(\M^u\right)-|\det \left(\M^u\right)|\left(|M^u_{33}|\M^u_{12}n^{d}_{2}+|\M^u_{22}|M^d_{33}n^{u}_{1}\frac{\delta_u}{\delta_d} \right)^\ast \Big]\, ,
\end{align}
which are generically $\mathcal{O}(1)$ quantities, we have 
\begin{align}
J=\Bigg[\frac{M_{33}^d \M_{22}^u \M_{22}^{d\ast}n^{u}_{2}}{|\M_{22}^u|^3 |\det \left(\M^u\right)\M^d_{22} M_{33}^d M_{33}^u|^2}-\frac{|M_{33}^u| \M_{22}^{d\ast}n^{d}_{2}}{|\M_{22}^u|^2 |\det \left(\M^u\right)\M^d_{22} M_{33}^d M_{33}^u|^2}\frac{\delta_d}{\delta_u}  \Bigg]\mathcal{J}_1 \mathcal{J}_2 \textcolor{red}{\delta_u \delta_d} \textcolor{blue}{\epsilon_L^2}
\end{align}

\section{Basis of $Sp(4)$ generators}

For ease of reference, a basis for the 10-dimensional Lie algebra $\mathfrak{sp}(4)$, in the defining representation, is
\begin{align}
&\lambda_1 = \frac{1}{2}\begin{pmatrix}
1&0&0&0\\
0&0&0&0\\
0&0&-1&0\\
0&0&0&0
\end{pmatrix}, \qquad
&&\lambda_2 = \frac{1}{2}\begin{pmatrix}
0&0&0&0\\
0&1&0&0\\
0&0&0&0\\
0&0&0&-1
\end{pmatrix}, \\
&\lambda_3 = \frac{1}{2\sqrt{2}}\begin{pmatrix}
0&1&0&0\\
1&0&0&0\\
0&0&0&-1\\
0&0&-1&0
\end{pmatrix}, \qquad
&&\lambda_4 = \frac{1}{2\sqrt{2}}\begin{pmatrix}
0&i&0&0\\
-i&0&0&0\\
0&0&0&i\\
0&0&-i&0
\end{pmatrix},\\
&\lambda_5 = \frac{1}{2}\begin{pmatrix}
0&0&1&0\\
0&0&0&0\\
1&0&0&0\\
0&0&0&0
\end{pmatrix}, \qquad
&&\lambda_6 = \frac{1}{2}\begin{pmatrix}
0&0&i&0\\
0&0&0&0\\
-i&0&0&0\\
0&0&0&0
\end{pmatrix},\\
&\lambda_7 = \frac{1}{2}\begin{pmatrix}
0&0&0&0\\
0&0&0&1\\
0&0&0&0\\
0&1&0&0
\end{pmatrix}, \qquad
&&\lambda_8 = \frac{1}{2}\begin{pmatrix}
0&0&0&0\\
0&0&0&i\\
0&0&0&0\\
0&-i&0&0\end{pmatrix},\\
&\lambda_9 = \frac{1}{2\sqrt{2}}\begin{pmatrix}
0&0&0&1\\
0&0&1&0\\
0&1&0&0\\
1&0&0&0
\end{pmatrix}, \qquad
&&\lambda_{10} = \frac{1}{2\sqrt{2}}\begin{pmatrix}
0&0&0&i\\
0&0&i&0\\
0&-i&0&0\\
-i&0&0&0
\end{pmatrix}
\end{align}
The normalization is such that
\begin{equation}
\Tr (\lambda_a \lambda_b) = \frac{1}{2} \delta_{ab}\, .
\end{equation}

\bibliographystyle{JHEP}
\bibliography{Sp4}
\end{document}